\documentclass[letterpaper,11pt]{article}
\usepackage{authblk}
\usepackage{amssymb}
\usepackage{tabularx}
\usepackage{subcaption}
\usepackage{multirow}
\usepackage{amsmath}
\usepackage{graphicx}
\usepackage[margin=1in,letterpaper]{geometry} 
\usepackage{cite}
\usepackage[final]{hyperref}
\usepackage[dvipsnames]{xcolor}
\usepackage{listings}
\usepackage[toc,page]{appendix}
\usepackage{docmute}

\makeatletter
\newcommand*{\rom}[1]{\expandafter\@slowromancap\romannumeral #1@}
\makeatother
\bibliography{journals,phd-references} 
\graphicspath{{images/}}
\DeclareGraphicsExtensions{.jpg}


\hypersetup{
	colorlinks=true,       
	linkcolor=blue,        
	citecolor=blue,        
	filecolor=magenta,     
	urlcolor=blue         
}

\begin{document}

\title{Finding the origin of noise transients in LIGO data with machine learning}
\author[1]{Marco Cavagli\`{a}}
\author[2]{Kai Staats}
\author[2]{Teerth Gill}
\affil[1]{Department of Physics and Astronomy, The University of Mississippi\break University MS 38677-1848, USA}
\affil[2]{Department of Physics and Astronomy, Embry-Riddle University\break Prescott AZ 86301, USA}
\date{\today}
\maketitle
\begin{abstract}

\noindent Quality improvement of interferometric data collected by gravitational-wave detectors such as Advanced LIGO and Virgo is mission
critical for the success of gravitational-wave astrophysics. Gravitational-wave detectors are sensitive to a variety of disturbances of
non-astrophysical origin with characteristic frequencies in the instrument band of sensitivity. Removing non-astrophysical artifacts that
corrupt the data stream is crucial for increasing the number and statistical significance of gravitational-wave detections and enabling
refined astrophysical interpretations of the data. Machine learning has proved to be a powerful tool for analysis of massive quantities of
complex data in astronomy and related fields of study. We present two machine learning methods, based on random forest and genetic
programming  algorithms, that can be used to determine the origin of non-astrophysical transients in the LIGO detectors. We use two classes of
transients with known instrumental origin that were identified during the first observing run of Advanced LIGO to show that the algorithms can
successfully identify the origin of non-astrophysical transients in real interferometric data and thus assist in the mitigation of
instrumental and environmental disturbances in gravitational-wave searches. While the datasets described in this paper are specific to LIGO,
and the exact procedures employed were unique to the same, the random forest and genetic programming code bases and means by which they were
applied as a dual machine learning approach are completely portable to any number of instruments in which noise is believed to be generated
through mechanical couplings, the source of which is not yet discovered.

\end{abstract}

\section{Introduction}\label{Introduction}

\noindent On February 11$^{\rm th}$, 2016, scientists from the Laser Interferometer Gravitational-wave Observatory (LIGO)
\cite{TheLIGOScientific:2014jea} Scientific Collaboration (LSC) and the European Virgo Collaboration \cite{Acernese:2015gua} announced the
first direct detection of gravitational waves from a coalescing pair of two stellar-mass black holes \cite{Abbott:2016blz}. Detection of the
GW150914 gravitational-wave signal, recorded at the LIGO sites in the morning of September 14$^{\rm th}$ 2015, marks the beginning of a new
observational era in astrophysics. Strong, dynamical relativistic gravitational fields can now be used to map the dark universe and probe
fundamental physics. One hundred years after the formulation of general relativity, fifty years after the pioneering work of Joseph Weber, and
several decades after the foundation of LIGO, gravitational-wave astrophysics is a reality. 

The next decade will see this new branch of scientific research expand to a mature field \cite{natacad}. Since GW150914, another four
gravitational-wave detections from binary black hole systems \cite{Abbott:2016nmj,Abbott:2017vtc,Abbott:2017oio,Abbott:2017gyy} and a detection
from a binary neutron star system \cite{TheLIGOScientific:2017qsa} were recorded in the data stream of the Advanced LIGO and Virgo
interferometers. More varied detections are anticipated in future LIGO and Virgo observation runs
\cite{Abbott:2016ymx,Abbott:2016nhf,Aasi:2013wya}, spurring a plethora of astrophysical and theoretical investigations. KAGRA
\cite{Akutsu:2015hua} and LIGO-India \cite{Unnikrishnan:2013qwa} will join the international network, enormously improving localization of
astrophysical sources and the network duty cycle. Commissioning activities will strive to bring the instruments to design sensitivity.
Instrumental R\&D will focus on the design and realization of the next generation of gravitational-wave interferometric detectors on Earth
\cite{Evans:2016mbw} and in space \cite{lisa}. All these activities will be crucial for the growth of gravitational-wave astrophysics from a
sensational news item to a full-grown scientific method to explore our universe. 

The measured rate of gravitational-wave detections in the first observing run (O1) and second observing run (O2) of Advanced LIGO and Virgo
implies that the international network of interferometers is poised to detect a significant number of gravitational-wave events in the coming
years. The third Advanced LIGO-Virgo observing run (O3) is scheduled for early 2019. As the gravitational-wave detector network reaches a stage
that supports rates of detections of astrophysical gravitational-wave sources as high as $\sim 1~{\rm day}^{-1}$, a fast and accurate
assessment of data quality will be critical.

The Advanced LIGO and Virgo detectors are sensitive to a variety of disturbances of non-astrophysical origin with characteristic frequencies in
the instrument band of sensitivity \cite{Slutsky:2010ff,TheLIGOScientific:2016zmo}. Noise transients of instrumental or environmental origin
increase the false alarm rate of searches for gravitational-wave bursts and compact binary coalescences as well as affect measurements of these
signals. The most remarkable example of the effect of a noise transient on a gravitational-wave signal is undoubtely the glitch that occurred in
the LIGO-Livingston detector in coincidence with the binary neutron star merger detection \cite{TheLIGOScientific:2017qsa} and had to be
carefully modeled and subtracted from the data to accurately determine the properties of the signal. Noise in the frequency domain affects
searches for long-lived transients, continuous waves and stochastic background. Removing non-astrophysical artifacts from the data and improving
the background of LIGO's searches is crucial for reducing non-stationarity in the detectors, extending the network duty cycle, and increasing
the statistical significance of gravitational-wave candidate events. Improvements in these areas, in turn, boost parameter estimation of the
signals and enable refined astrophysical interpretations of the data. For all of these reasons, the understanding and mitigation of
non-astrophysical disturbances in the detectors is one of the top priorities of the LSC and the Virgo Collaboration.

In recent years, a significant part of LSC and Virgo activities has been devoted to investigations aimed at characterizing non-astrophysical
noise, improving data quality of gravitational-wave searches and detector commissioning. Examples of these activities include investigations
of noise transients and spectral features of known and unknown origin, studies of correlations between environmental and instrumental
channels, detector performance assessment, and generation of data quality flags and vetoes for LIGO and Virgo's searches. Many of these
activities are conducted by instrument specialists, commissioners and data analysts working together to identify, categorize and mitigate
undesirable noise transients and spectral features that corrupt LIGO-Virgo gravitational-wave searches. These tasks are generally performed by
mining the data of the gravitational-wave strain channel and a large number ($\sim$ 400,000) of environmental and instrumental auxiliary
channels. Data quality investigations and detector characterization were critical to validating the first detections by reducing
gravitational-wave search backgrounds and increasing the significance of the signals. 

Over the years, LSC researchers have developed and implemented many methods and algorithms to perform these tasks
\cite{Slutsky:2010ff,TheLIGOScientific:2016zmo}. Some of the data quality tools run {\it online} to provide a fast assessment of the status of
the interferometers for low-latency searches. A different set of tools is used {\it offline} for deeper searches and follow-up of
gravitational-wave candidates. With the LIGO detectors striving to reach design sensitivity and an anticipated large number of
gravitational-wave detections in the upcoming observing runs, understanding and mitigating instrumental and environmental noise sources will
become increasingly more important. 

The expected increase in detections from compact binary systems and the discovery of gravitational waves from other kinds of astrophysical
sources will likely render current methods for data quality assessment inadequate for the tasks ahead. For this reason, the development of
improved methods to investigate noise and the exploration of new approaches to data quality issues are recognized priorities of LIGO and Virgo
researchers.

Ground-based interferometric gravitational-wave detectors are complex devices that exhibit non-linear couplings across instrumental subsystems
and the environment. As a consequence, the LIGO-Virgo  detector noise is non-Gaussian, variable across many parameters, and cannot be fully
analytically modelled. Machine Learning (ML) customarily denotes the science of design, development, and  applications of computer algorithms
that ``learn'' to perform specific tasks and automatically improve their performance through the use of adaptive techniques and iterative
procedures. Methods based on computational learning theory are powerful tools
to analyse complex system data and may prove valuable for improving the manner in which LIGO and Virgo operate in the area of data quality.

Recently, several groups in the LSC and Virgo collaboration have investigated the use of ML techniques for data analysis and detector
characterization. The ``Gravity Spy'' project, for example,  aims at using citizen science and ML for classification of LIGO noise transients
over the next observing runs \cite{Zevin:2016qwy}. Supervised deep learning algorithms have been proposed for  glitch classification
\cite{Mukund:2016thr, CuocoRazzano, George:2017qtr,George:2018awu} as well as real-time gravitational-wave detection \cite{George:2016hay},
parameter estimation \cite{George:2017pmj} and signal classification  \cite{Kapadia:2017fhb, Vinciguerra:2017psh,Gabbard:2017lja,Gebhard:2017}.
Multivariate random forest classifiers \cite{iDQ,Biswas:2013wfa} and unsupervised ML algorithms based on Principal Component Analysis
\cite{Powell:2015ona,Powell:2016rkl} have been used on interferometric data over the years. ML methods can provide complementary approaches to
existing detector characterization techniques as they  are computationally inexpensive, able to deliver results in low latency, and derive
predictive models of the system producing the data.

In this paper we present a {\it new} application of ML to an {\it old} problem in experimental detection physics: the identification of
instrumental mechanical couplings leading to excess noise in the detector. Our analysis focuses on ground-based gravitational-wave
interferometric detectors, in particular the LIGO-Virgo instruments. However, the methods presented here are general and can be applied to any
complex physical device with a main output channel and a set of auxiliary channels that monitor the status of the instrument and its
environment. 

The ultimate goal of detector characterization is not only to flag and veto noisy times in the instrument main output data stream, but also 
identify the instrumental or environmental source of the noise and, if possible, adjust the detector such that the disturbance can be removed
permanently. Whereas different ML techniques have been developed by various LSC groups to classify noise transients, or glitches, ML has not
been applied yet to the problem of identifying the cause of non-astrophysical noise in LIGO-Virgo detectors. Herein lies the novelty of our
approach: We introduce two ML algorithms that provide simple, yet robust methods to mine the data of auxiliary channels and infer the origin of
noise transients in the  main detector output. The codes developed for this task are based on two, widely-used flavors of ML known as Random
Forest (RF) and Genetic Programming (GP). This choice is motivated by our final  goal of providing fast and effective tools that commissioners
and data analysts can use with little tuning. Contrarily to more black box approaches to ML such as deep learning and neural network-based
algorithms, RF and GP methods are interpretable, easy to use and tune, and can work with relatively small datasets without the inherent risk of
overfitting. The methods that we illustrate below only require an input list of times when a specific class of noise transients occurs. Rather
than generating ML features in the form of time-frequency images for deep learning image-based classification (a time-consuming process), the
required features are drawn directly from numerical metadata that are generated by real-time data quality pipelines for generic detector
characterization investigations readily already available on the LIGO computing clusters. This approach minimizes the feature generation step of
the process, which is the typical bottleneck for low-latency investigations. The ML dataset is readily assembled by putting together the
features of the noise transients and additional, randomly-selected background triggers. The RF and GP codes that we developed can be trained and
run in minutes on LIGO computing clusters and allow the user to complete a typical analysis with a number of noise transients of the order of a
few thousands in low-latency and with minimal input.

In the following sections we illustrate this approach by applying the RF and GP methods to two set of glitches in LIGO data from LIGO-Virgo O1
and O2 observing runs. The origin of these  glitches was identified by LIGO-Virgo commissioners and scientists, eventualy leading to the
successful mitigation of these disturbances and their removal from later data. While our present work  does not have a direct impact on the
searches for transient gravitational waves with O1 and O2 data, it presents a proof-of-concept that our method can be implemented within the
current LIGO-Virgo  data quality infrastructure, which will be in place in the upcoming observing runs, and used to infer instrumental and
environmental mechanical couplings affecting the detectors in O3 and beyond.

\section{Machine learning algorithms}\label{Algos}

\noindent In this section we briefly introduce the basics of the RF and GP algorithms. A full introduction to these methods is beyond the
scope of this paper and we only present information which is essential for the understanding of our analysis. For a deeper discussion of RF
and GP, the reader is referred to Refs.\ \cite{RFweb, GPweb}.

\subsection{Random Forests}\label{RF}

\begin{figure}[htbp]
  \begin{center}
   \includegraphics[width=140mm]{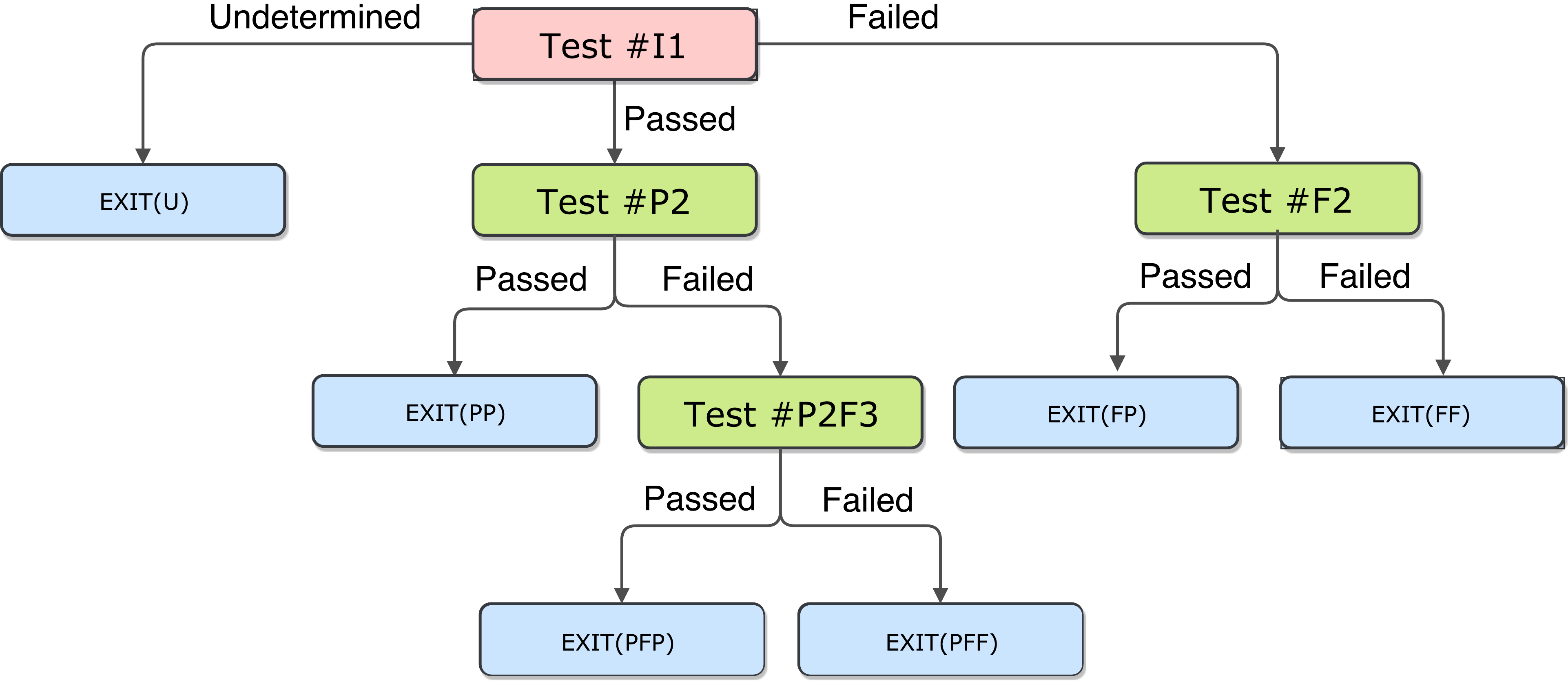}
  \end{center}
  \caption{Graphical representation of a decision tree. The root node is represented by the red element, internal nodes are represented by green elements, leaves are represented by blue elements and branches are represented by black connector lines which denote success or failure of the stated condition.}\label{fig:decision_tree}
\end{figure}

\noindent RF denote a popular supervised ML technique for data classification and regression which operates without human intervention,
employing off-line training (learning) and test (validation) to produce the outcome. The basic principle of the RF method is the construction
of a number of decision trees at training time and then averaging over these trees to improve their perfomance on the testing set. A decision
tree can be represented by a graph with three types of elements representing tests on input features (internal nodes), test outcomes
(branches) and labels that are used to make predictions (leaves). The topmost node in a tree is called the root node. Paths from root to end
leaves represent classification or regression rules. An example of decision tree is shown in Fig.\ \ref{fig:decision_tree}.

Single decision trees may have high bias and overfit the training set,  as well as be very sensitive to outlier data. In order to avoid this,
the RF method calculates a combination of independently-sampled tree predictors. This averaging procedure avoids both false minima and
over-training leading to a single, more complex, final tree which is less likely to overfit. One of the main advantages of RF is the
transparent nature of its computational algorithm. Ensemble algorithms such as RF are typically more powerful than other ML techniques. They
require little data preparation and are easy to interpret. As they have few hyperparameters to tune, they can produce accurate results even
with default settings. Feature importance and algorithm accuracy are generated automatically.

In our analysis we used the standard scikit-learn \cite{RFweb} implementation of the RF algorithm with default settings with the exception of
the number of estimators (trees in the forest) that we chose in the range of 300-500. The RF methods allows for a straightforward computation
of the feature importance, i.e.,  how much each feature contributes to the model's predictive performance. In the scikit-learn implementation
of the RF algorithm, the feature importance is defined as the ``gini importance'' \cite{Breiman1984}, i.e., the average over all trees of the
total decrease in node impurity, weighted by the probability of reaching that node.

In order to increase the signal-to-noise ratio of the RF algorithm, we refit multiple times with each iteration removing features below a
user-defined threshold until no features below the chosen cutoff value are left. This process allows us to denoise the output by eliminating
features that are marginal in the determination of the source of the glitches. The features belonging to the same auxiliary channel are then
grouped together to build the channel importance, which measures the relevance of the given channel in discriminating generic background noise
triggers (label 0) from the glitches inder investigation (label 1). 

\subsection{Genetic programming}\label{GP}

\noindent GP is a supervised machine learning algorithm, an analog to biological natural selection that evolves a population of programs to
solve a particular problem \cite{Koza1992}. An individual GP program is an hypothesis which when executed takes the form of a mathematical,
multivariate expression.

In training, GP compares the output of each executed hypothesis (predict) to an associated, qualified label (truth). This comparison is
quantified as a fitness score. GP programs that demonstrate a higher fitness score are more likely (but not guaranteed) to be selected for the
next generation. Thus, each subsequent generation of programs is more likely (but not guaranteed) to solve the given problem than the prior
\cite{Poli2008}.

\begin{figure}[htbp]
  \begin{center}
    \begin{tabular}{  l  l  } 
    \includegraphics[width=120mm]{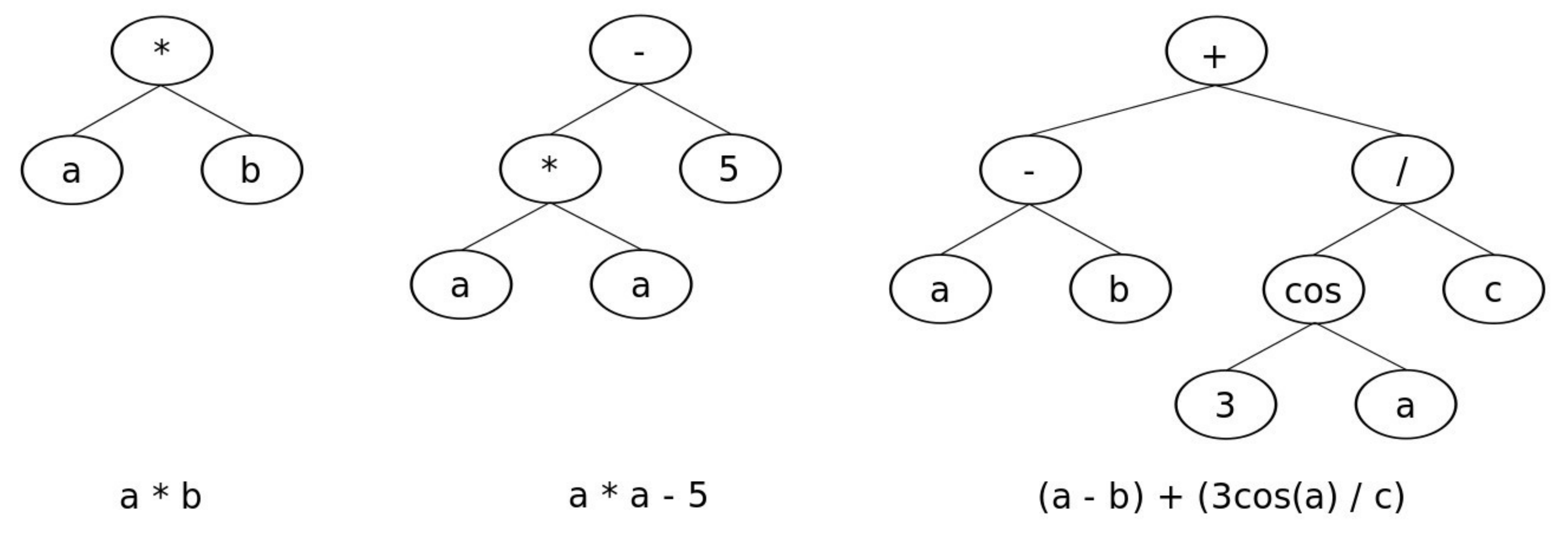}
   \end{tabular}
  \end{center}
  \caption{Graphical representation of a GP syntax tree. The depth of a tree is defined as the number of rows in the tree, i.e., two, three
  and four in the diagrams from left to right, respectively.}\label{fig:genetic_trees}
\end{figure}

As shown in Figure \ref{fig:genetic_trees}, GP multivariate expressions are often represented as a syntax tree, where the trees have a root
(top center), nodes (mathematical operators), and leaves (operands). Operators can be arithmetic, trigonometric, and boolean, for example.
Operands are variable place-holders for the real-world data. When evaluated, the real-world values of the data are substituted for the
variables in the multivariate expressions, data point by data point. The depth of a tree determines the complexity of the resulting, evolved
multivariate expression. 

User defined parameters affect the quality and speed of the evolutionary process, including the size and type of GP trees in the initial
population, the number of GP programs selected for each fitness score comparison and the type of comparison applied, and the termination
criterion (eg: number of generations).

\begin{figure}[htbp]
  \begin{center}
    \begin{tabular}{  l  l  } 
    \fbox{\includegraphics[width=60mm]{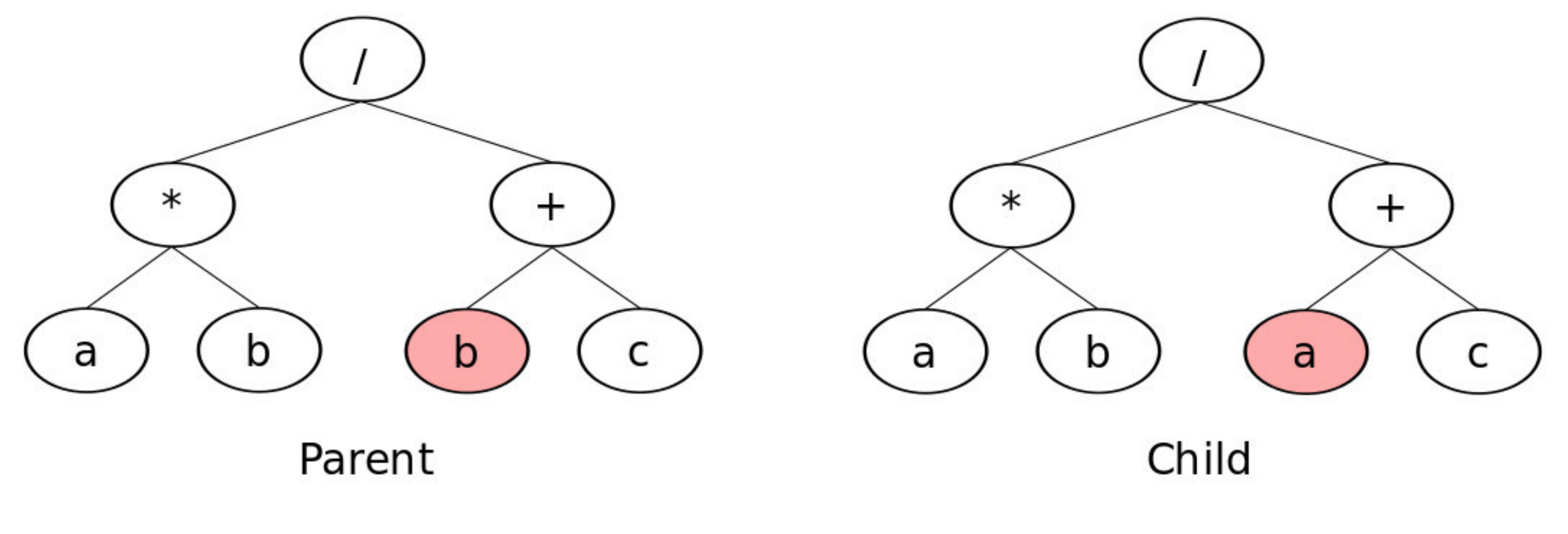}}&\multirow{-4}{*}{\fbox{\includegraphics[width=75mm]{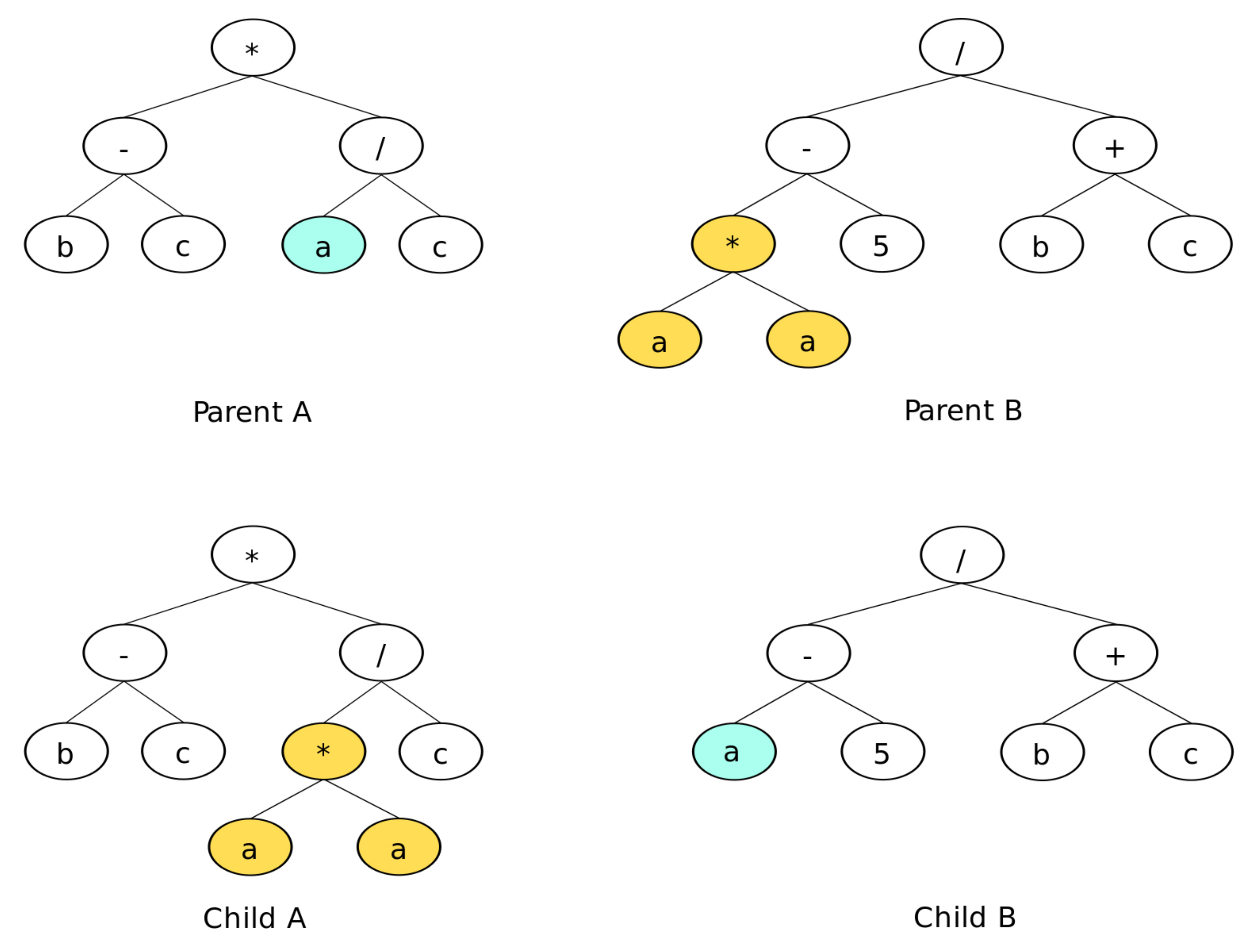}}}\\&\\
    \fbox{\includegraphics[width=60mm]{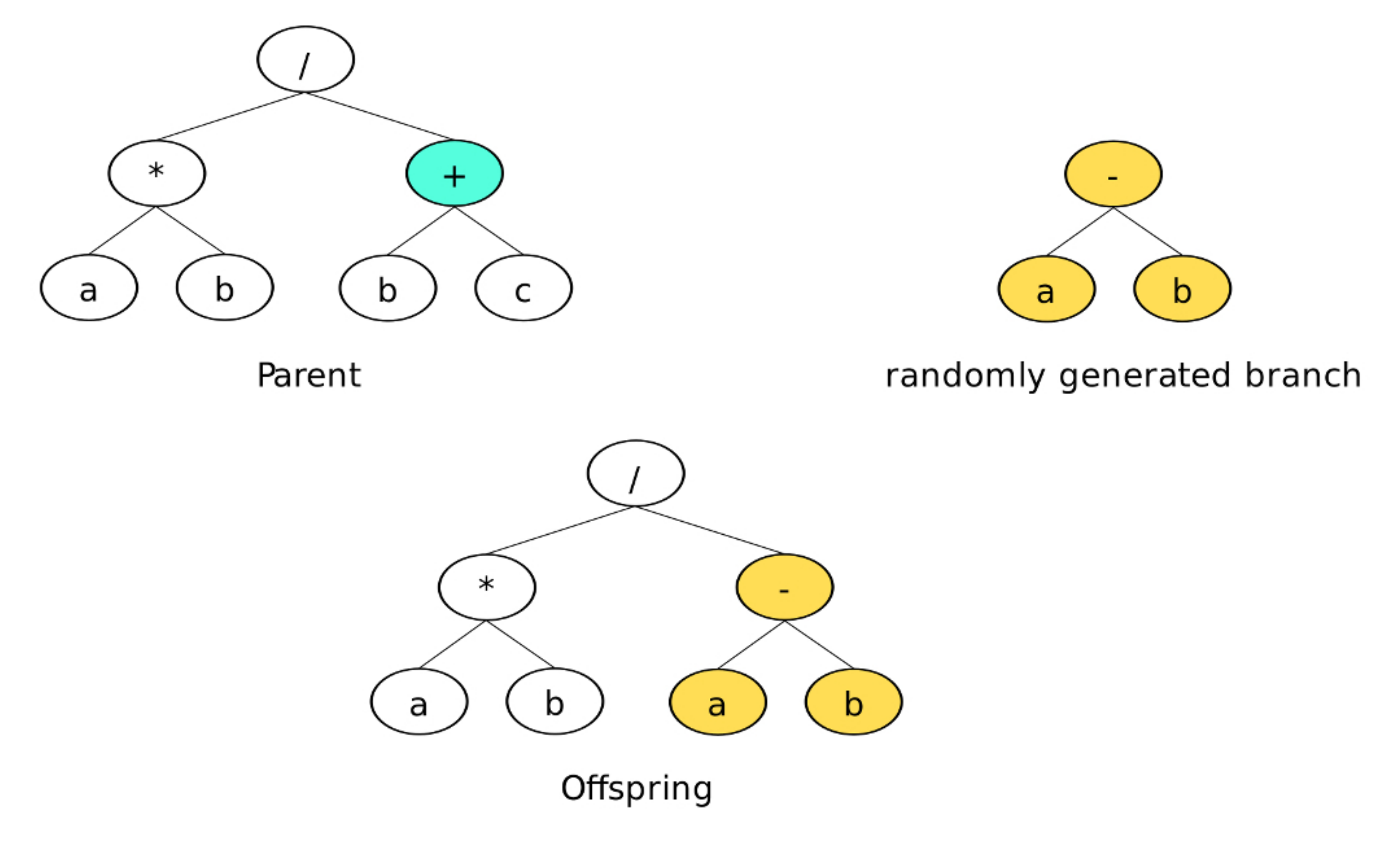}}&\\
   \end{tabular}
  \end{center}
  \caption{Genetic operators for evolutionary computation: Point mutation (top left), branch mutation (bottom left), and crossover or sexual
reproduction (right). Reproduction is defined as no change in a tree when copied from the current to the next generation.}
  \label{fig:genetic_operators}
\end{figure}

The work-flow of a generational GP run incorporates three basic steps: a) Generation of an initial, stochastic population; b) Iterative
selection, evaluation, and application of genetic operations (reproduction, mutation and crossover --see Fig.\ \ref{fig:genetic_operators});
c) transfer of the evolved copy into the subsequent generation. Steps b) and c) are repeated until the user-defined termination criteria are
met \cite{Poli2008}.

Key to the acceptance of GP across many fields of research is the transparent nature of its computational engine. At any stage of the
evolutionary process, the internal workings of GP can be readily exposed and reviewed, and the populations archived. As compared to other,
more black box machine learning algorithms, GP provides insight to how it arrives to its evolved solution. Moreover, as the GP model is a
stand-alone mathematical expression whose variables call upon data features, it can be readily employed as a portable model for online data
classification or regression analysis. The algorithm can be tuned by choosing a number of user-defined hyperparameters, which include base,
maximum and minimum tree depth, the size of the program population, the number of generations and the tournament size. 

In our analysis we used a tree-based open source python code, Karoo GP \cite{KarooGP}, that was originally written by one of the authors (KS)
for the mitigation of RFI in radio astronomy at the Square Kilometre Array \cite{StaatsThesis}. Karoo GP is scalable, with multicore and GPU
support enabled by the library TensorFlow, with capacity to work with very large datasets \cite{Gecco2017}. 

\section{Data sets for algorithm testing}\label{datasets}

\noindent In order to illustrate how ML can be used to identify non-astrophysical transients in LIGO data and recover mechanical couplings in
the detector subsystems, we consider two sets of glitches with known origin in analysis-ready data from Advanced LIGO's first Observing Run
(O1: September 12$^{\rm th}$, 2015, to January 19$^{\rm th}$, 2016) and second Observing Run (O2: November 30$^{\rm th}$, 2016, to August
25$^{\rm th}$, 2017). Here ``analysis ready'' denotes data taken with the interferometers in a nominal observing state. 

During observing runs the status of the LIGO interferometers is continuously monitored through a number of physical sensors that probe the
detector subsystems and their environment. The digital output of these sensors is recorded in thousands of auxiliary channels as raw time
series. Dedicated detector characterization pipelines use these raw data to identify non-astrophysical noise transients and spectral features
that may affect the instrument main output, or gravitational-wave strain channel. Auxiliary channels are typically separated in ``safe'' and
``unsafe'' channels. Safe auxiliary channels are not expected to show any excess noise when an astrophysical signal is present in the
gravitational strain channel. For example, physical environmental monitor (PEM) channels are considered safe, as a gravitational wave is not
supposed to generate any signal in environmental sensors. Thus excess noise in these channels generally denotes a non-astrophysical
disturbance.

Data quality flags and ultimately vetoes can be created from safe channels to reduce the instrumental background and improve the
gravitational-wave searches. Unsafe auxiliary channels are known to couple to astrophysical signals. If a gravitational wave is present in the
data, its signal is expected to couple to some of the detector subsystems, for example the interferometer output mode cleaner. Unsafe channels
cannot be used to create vetoes and are not considered in detector characterization investigations. Standard lists of safe and unsafe channels
are made for each observing run and periodically tested through insertions of simulated signals in the instrument (hardware injections). In
our analysis we consider the standard O1 and O2 lists of safe auxiliary channels as determined by the detector charaterization working group
and used by the hveto pipeline \cite{Smith:2011an}, comprising 840 and 919 channels, respectively.

The first set that we consider in our study contains 2049 glitches that were identified by the hveto pipeline in a magnetometer located at one
of the end-test stations of the LIGO-Livingston interferometer between February 9$^{\rm th}$, 2017 and April 10$^{\rm th}$, 2017. LIGO's
equipment which is hosted in electronics racks generates magnetic fields that may couple to other components of the detector, such as cables,
connectors and actuators. In order to measure these spurious magnetic fields, several magnetometers are deployed in the main interferometer
stations. These magnetometers are also used to monitor DC power supply glitches and currents that may produce artifacts in the gravitational
wave channel \cite{T1200221}. 

After a power outage on February 7$^{\rm th}$, 2017 the hveto pipeline identified a series of new glitches of electromagnetic origin in the
LIGO-Livingston interferometer within the detector search frequency band, around $\sim$ 50-60 Hz. The noise transients appeared as short-lived
spikes in one of the PEM-EX\_MAINSMON\_EBAY\_DQ auxiliary channels, the Physical Environmental Monitor (PEM) mains voltage monitor (MAINSMON)
of the Electronics Bay (EBAY) in the X-arm end station (EX), as well as in EX magnetometers. In the days following their first appearance,
these glitches amounted to about 15\% of all short-lived noise transients seen in the gravitational-wave strain channel. After a few weeks of
testing and diagnostics by the LSC detector characterization group, on April 11$^{\rm th}$ LIGO commissioners successfully mitigated the
glitches by modifying the grounding system of the electronics bay with the installation of a ground rod which eliminated a spurious current at
the origin of the electromagnetic disturbance.

The EX magnetometer glitches provide a good playground for ML testing because of their distinct spatial and temporal localization, as well as
their understood origin and successful mitigation. Moreover, due to their electromagnetic nature they appear only in a well-defined subset of
auxiliary channels. This allows for a clear-cut test of the algorithm's ability to identify the correct auxiliary channels related to the
noise and infer the source of the instrumental coupling.

The second set contains 42 short-lived noise transients that were caused by an air compressor seismically coupling to the LIGO-Hanford
interferometer between September 18$^{\rm th}$, 2015 and December 28$^{\rm th}$, 2015. In September 2015 detector characterization
investigations indicated the presence of some transient excess noise of unknown origin at a frequency around 50 Hz in the gravitational strain
output of the LIGO-Hanford interferometer. The origin of the noise was recognized to originate in the EX station. The hveto pipeline indicated
a correlation in time with glitches in the EX PEM seismic (SEIS) and accelerometer (ACC) auxiliary channels, as well as the Streckeisen STS-2
(STS) inertial sensors monitoring the ground motion (GND) of the active seismic isolation internal to the vacuum system (ISI). The effect of
the excess noise due to the seismic coupling on the transmitted light (TR) along the direction of the interferometer X arm caused some excess
pitch and yaw motion of the EX test mass that was recorded in the Alignment Sensing and Control (ASC) channels. Dedicated investigations
pointed to a mechanical coupling as the origin of the transients, such as a motor or a transformer core pulse inducing the 50 Hz
characteristic. A time delay between the occurrence of the noise in the accelerometers and the voltage monitors indicated that the origin of
the transients was not located in the immediate vicinity of the EX optics table. The culprit was eventually identified as an air compressor
turning on in the EX station which was seismically coupling to the detector via the optics table. Follow-up investigations with Gravity Spy
\cite{Zevin:2016qwy} led to the identification of all the 42 glitches in our dataset. Although the overall number of these glitches is quite
low for ML training purposes, their physical properties are well characterized and their instrumental mechanical couplings are well understood.
Similarly to the magnetometer set, the air compressor glitches provide a good playground for testing the efficiency of the ML algorithms in
determining the origin of noise disturbances.

These diverse sets allow us to test the effectiveness of the ML algorithms in two extreme cases that are typical of LIGO noise investigations,
where thousands or just a handful of glitches can be identified by detector characterization pipelines or manual data mining techniques.

\subsection{Data set preparation}\label{dataprep}

\noindent Short-lived noise transients are generally identified by a trigger in the form of a single GPS time or a time interval where the
disturbance has its peak. The peak of the glitch may be computed by simply recording the time where the value of a given auxiliary channel
time series passes a pre-defined threshold or through more refined methods, for example by defining a Signal-to-Noise Ratio (SNR) or looking
at correlations between channels and/or the main interferometer output. 

Our analysis is based on Omicron triggers. Omicron is a widely used LIGO pipeline to identify glitches in instrumental and environmental
auxiliary channels \cite{omicron}. The algorithm is based on a C++ burst-type Event Trigger Generator (ETG) which is itself based on the
Q-transform \cite{chatterji-thesis}, a modification of the standard short-time Fourier transform similar in construction to the continuous
wavelet transform. The channel time series is projected onto a parameter space tiled in time, frequency, and Q planes. An omicron trigger is
identified when the SNR of a time-frequency tile is above a user pre-defined threshold. The characteristics of each trigger are recorded in a
user-configurable data vector describing the physical parameters of the trigger. Elements of this data vector can be used as raw features for
the ML algorithm. In our analysis, we use a six-dimensional omicron data vector with peak frequency, central frequency, bandwidth, amplitude,
SNR and phase elements. 

The Omicron pipeline ran daily in O1 and O2 on the standard list of auxiliary channels, recording noise triggers with ${\rm SNR} > 5.5$. The
Omicron features corresponding to the magnetometer and air compressor glitches are obtained by selecting for each glitch time and auxiliary
channel the online Omicron trigger with the highest SNR within a coincidence window of $\pm 0.1$ seconds. As some of the auxiliary channels
may be not active at any given time during the observing run, we consider only the subset of triggers that have available Omicron triggers for
a minimum number of auxiliary channels. The higher the number of features and triggers in the data set, the better the ML training. However,
maximization of the number of features reduces the number of usable triggers in the set, as fewer Omicron triggers with all the required
auxiliary channels are available. A good trade-off between these two competing factors reduces the magnetometer set to 2000 triggers and 749
auxiliary channels, for a total of 4494 features per trigger, and the air compressor set to 16 triggers and 429 auxiliary channels, for a
total of 2856 features per trigger. The ML training datasets are completed by adding a number of background triggers (label 0) to the noise
triggers (label 1). These are obtained by randomly selecting a comparable number of Omicron triggers in the gravitational strain channel with
${\rm SNR} > 6$ that are not coincident with any of the glitch triggers (exclusion window equal to $\pm 10$ seconds) and then obtaining the
features for each auxiliary channel as done for the glitch set. Finally, label 0 and label 1 triggers in the datasets are randomized with 2/3
of the entries being used for ML training and internal validation, and the remaining 1/3 being reserved for testing.

\section{Results}\label{results}

\noindent In order to test the ML algorithms and illustrate how to infer the glitch mechanical couplings, we first ran the RF algorithm on the
training sets to compute the channel importance (defined as the sum of the feature importances for the given channel). We tested the procedure
by varying the number of estimators and the iteration threshold, noticing no significant change in the results. For illustration purposes,
here we present results for 500 estimators and an iteration threshold of 0.005 and 0 for the magnetometer and air compressor sets,
respectively. The RF results were successfully validated with Karoo GP, which was run multiple times with different hyperparameter
configurations to test the robustness of the procedure. Below we present results averaged on all Karoo GP runs, as well as some of the results
from the best performing runs.

\subsection{Magnetometer set}\label{magnetometerresults}

\begin{figure}[h]
  \begin{center}
   \includegraphics[width=160mm, height=100mm]{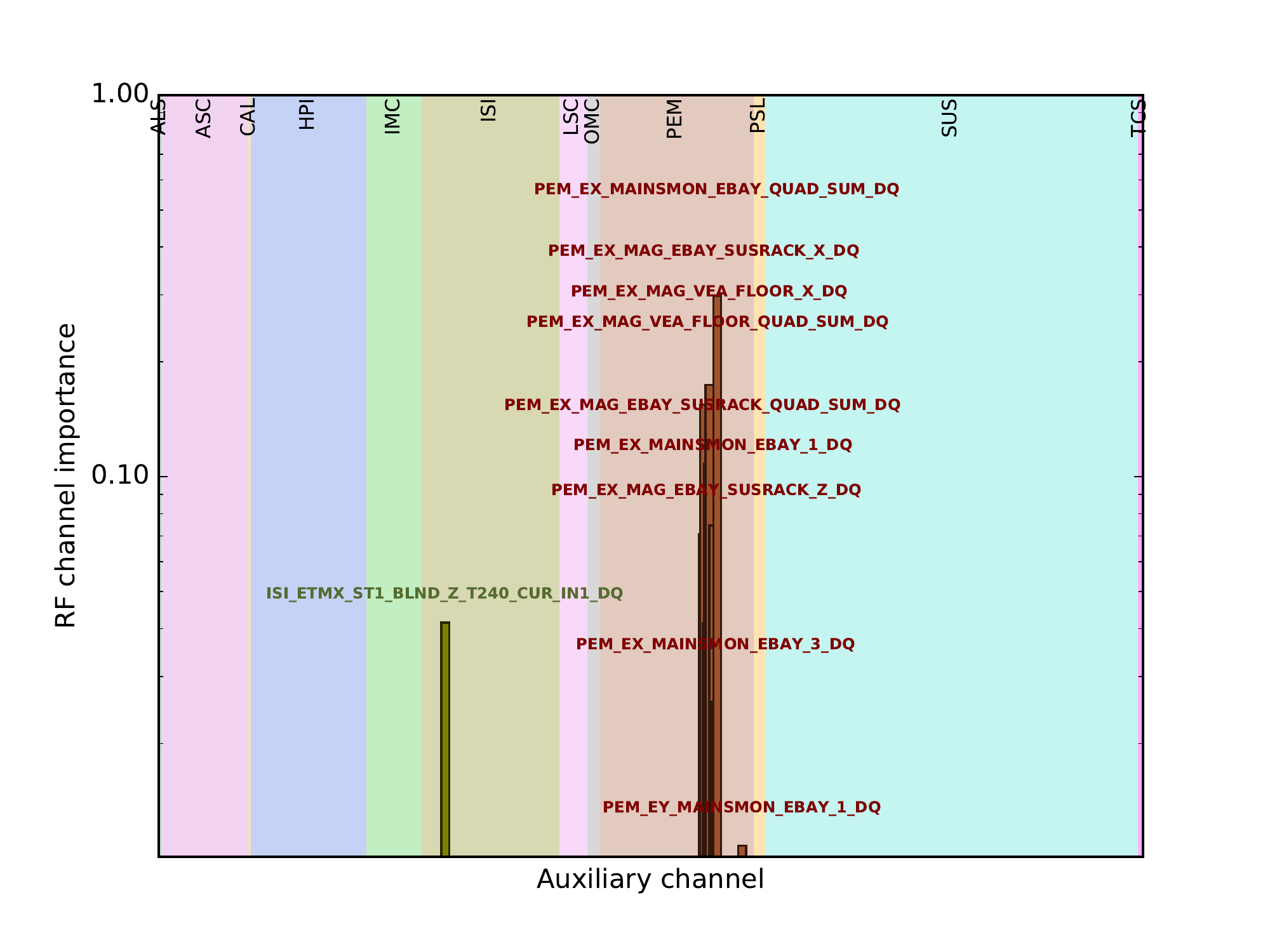}
  \end{center}
  \caption{Histogram of RF channel importance for the magnetometer set from the data in Table \ref{table:RF_magnetometer}. Different colors
denote different detector subsystems and auxiliary channels: Seagreen = Armlength Stabilization (ALS), orchid = Alignment Sensing and Control
(ASC), goldenrod = Photon Calibrator (CAL) royal blue = Hydraulic External Pre-Isolator (HPI), lime green = Input Mode Cleaner (IMC), olive
green = Internal Seismic Isolation (ISI), violet =  Length Sensing and Control (LSC), gray = Output Mode Cleaner (OMC), sienna = Physical and
Environmental Monitor (PEM), orange = Pre-Stabilized Laser (PSL), turquoise = Suspension (SUS), magenta = Thermal Compensation (TCS). The plot
clearly shows how the glitches arise from an environmental electromagnetic disturbance in the EX station.}\label{fig:RF_magnetometer}
\end{figure}

\noindent The list of auxiliary channels with nonzero importance obtained with RF iteration threshold equal to 0.005 is listed in Table
\ref{table:RF_magnetometer}. Out of the ten auxiliary channels, nine channels are related to the EX detector subsystem and one channel,
PEM\_EY\_MAINSMON\_EBAY\_1\_DQ, is the MAINSMON channel of the end-Y (EY) station's EBAY. As the latter has the lowest importance and can be
removed by choosing a higher RF iteration threshold while preserving most of the EX channels, we can safely conclude that it is not related to
the actual mechanical coupling originating the glitches. Eight of the auxiliary channels related to the EX subsystems are PEM channels for the
EBAY and the Vacuum Equipment Area (VEA) magnetometers and for the MAINSMON. The additional channel monitors the Nanometrics Trillium 240
(T240) Inertial broadband sensor picked off at the input to the BLEND filter bank of the active internal seismic isolation of the end test
mass. The data in Table \ref{table:RF_magnetometer} are shown as a histogram in Fig.\  \ref{fig:RF_magnetometer}.

\begin{table}
\begin{center}
  \begin{tabular}{  l  r  }
    {\bf Auxiliary channel} &{\bf RF Importance}\\ \hline\hline
    \color{OliveGreen} ISI-ETMX\_ST1\_BLND\_Z\_T240\_CUR\_IN1\_DQ&\color{OliveGreen} .041\\
    \color{RawSienna} PEM-EX\_MAG\_EBAY\_SUSRACK\_QUAD\_SUM\_DQ&\color{RawSienna} .071\\
    \color{RawSienna} PEM-EX\_MAG\_EBAY\_SUSRACK\_X\_DQ&\color{RawSienna} .155\\
    \color{RawSienna} PEM-EX\_MAG\_EBAY\_SUSRACK\_Z\_DQ&\color{RawSienna} .041\\
    \color{RawSienna} PEM-EX\_MAG\_VEA\_FLOOR\_QUAD\_SUM\_DQ&\color{RawSienna} .108\\
    \color{RawSienna} PEM-EX\_MAG\_VEA\_FLOOR\_X\_DQ&\color{RawSienna} .174\\
    \color{RawSienna} PEM-EX\_MAINSMON\_EBAY\_1\_DQ&\color{RawSienna} .075\\
    \color{RawSienna} PEM-EX\_MAINSMON\_EBAY\_3\_DQ&\color{RawSienna} .026\\
    \color{RawSienna} PEM-EX\_MAINSMON\_EBAY\_QUAD\_SUM\_DQ&\color{RawSienna} .298\\
    \color{RawSienna} PEM-EY\_MAINSMON\_EBAY\_1\_DQ&\color{RawSienna} .011\\
    \hline\hline
  \end{tabular}
  \caption{Auxiliary channels with nonzero RF importance for the magnetometer set. Different colors denote instrumental and environmental
auxiliary channels corresponding to different detector subsystems: Olive green = Internal Seismic Isolation (ISI), sienna = Physical and
Environmental Monitor (PEM).}\label{table:RF_magnetometer}
  \end{center}
\end{table}

The RF algorithm correctly identifies the voltage monitor of the EX electronics bay as the origin of the noise transients. The winning channel
is  PEM-EX\_MAINSMON\_EBAY\_QUAD\_SUM\_DQ, i.e., the quadrature sum of the raw EBAY MAINSMON output recorded by the Data Acquisition System
(DQ). The three MAINSMON channels account for about $\sim 40$\% of the features used in the RF classification. The PEM-EX\_MAG\_EBAY\_SUSRACK
channels recording the raw output of the EX rack magnetometer and the PEM-EX\_MAG\_VEA\_FLOOR channels recording the output of the VEA
magnetometer account for about $\sim 27$\% and $\sim 28$\% of the features, respectively. This result clearly points to an electromagnetic
origin of the glitches in the EX station, in agreement with the results first obtained with the hveto pipeline and later confirmed by LIGO
commissioners. The identification of the ISI-ETMX\_ST1\_BLND\_Z\_T240\_CUR\_IN1\_DQ by the RF code is interesting, as it may seem puzzling
that electromagnetic glitches show in the output of an accelerometer. A possible explanation for the inclusion of this channel could simply be
algorithm noise, as is the case with the PEM\_EY\_MAINSMON\_EBAY\_1\_DQ channel. However, the fact that this auxiliary channel is one of the
accelerometer channels in the EX station and cannot be eliminated by increasing the iteration threshold without also eliminating most of the
other EX channels suggests that the coupling may be real. Indeed,  broadband seismometers may couple to environmental magnetic fields
\cite{Forbriger2007,Forbriger2010}. In particular, the coherence of the T240 has been studied in detail in Ref.\ \cite{P1400087}, where it is
shown that strong magnetic fields may even dominate the seismometer signal. Thus the inclusion of ISI-ETMX\_ST1\_BLND\_Z\_T240\_CUR\_IN1\_DQ
indicates that the electromagnetic disturbance is sufficiently strong to couple to the ETMX seismometer, a fact that had not been recognized
during standard detector characterization investigations leading to the mitigation of the glitches.

\begin{figure}[h]
  \begin{center}
   \includegraphics[width=80mm]{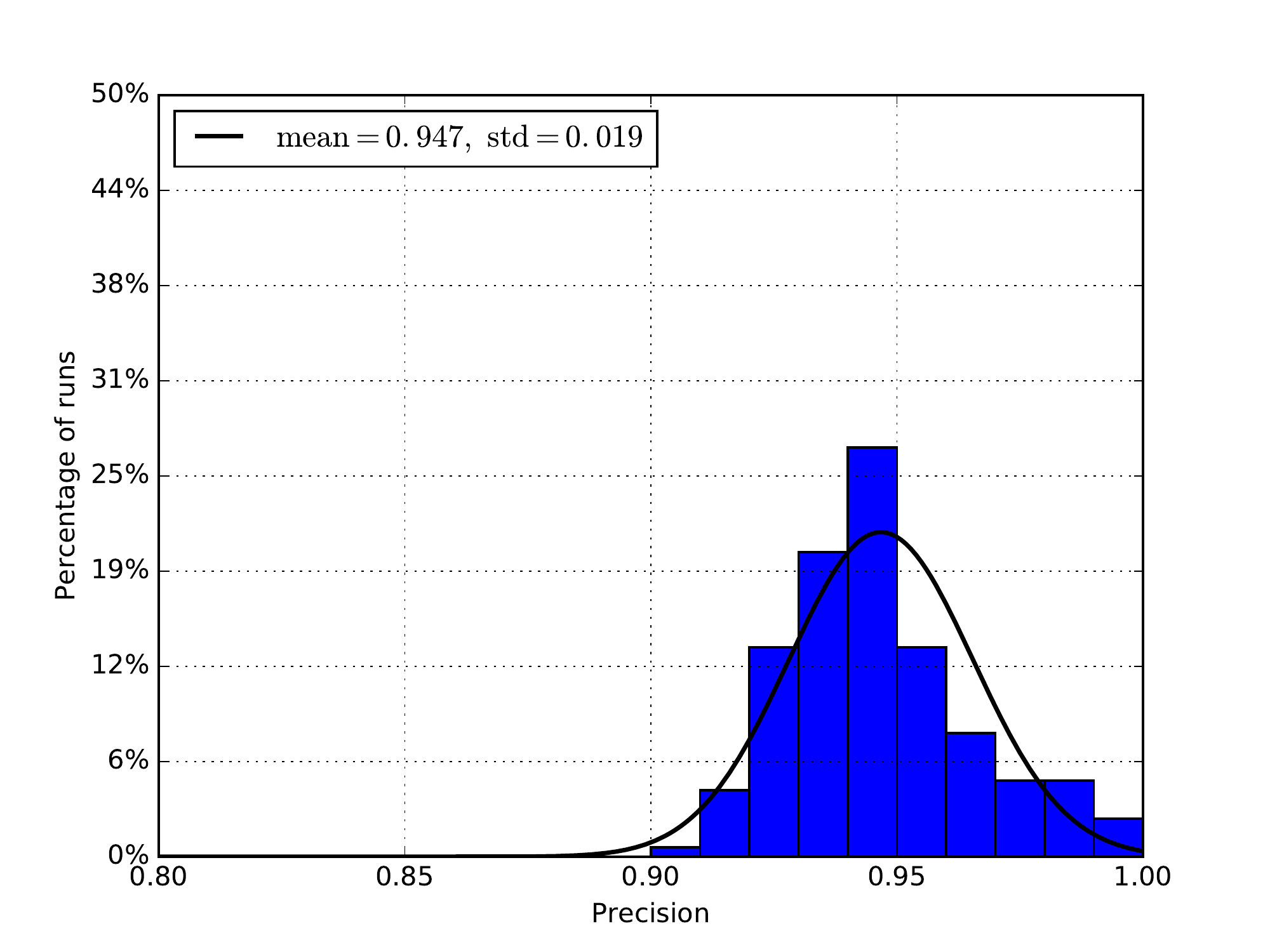}~\includegraphics[width=80mm]{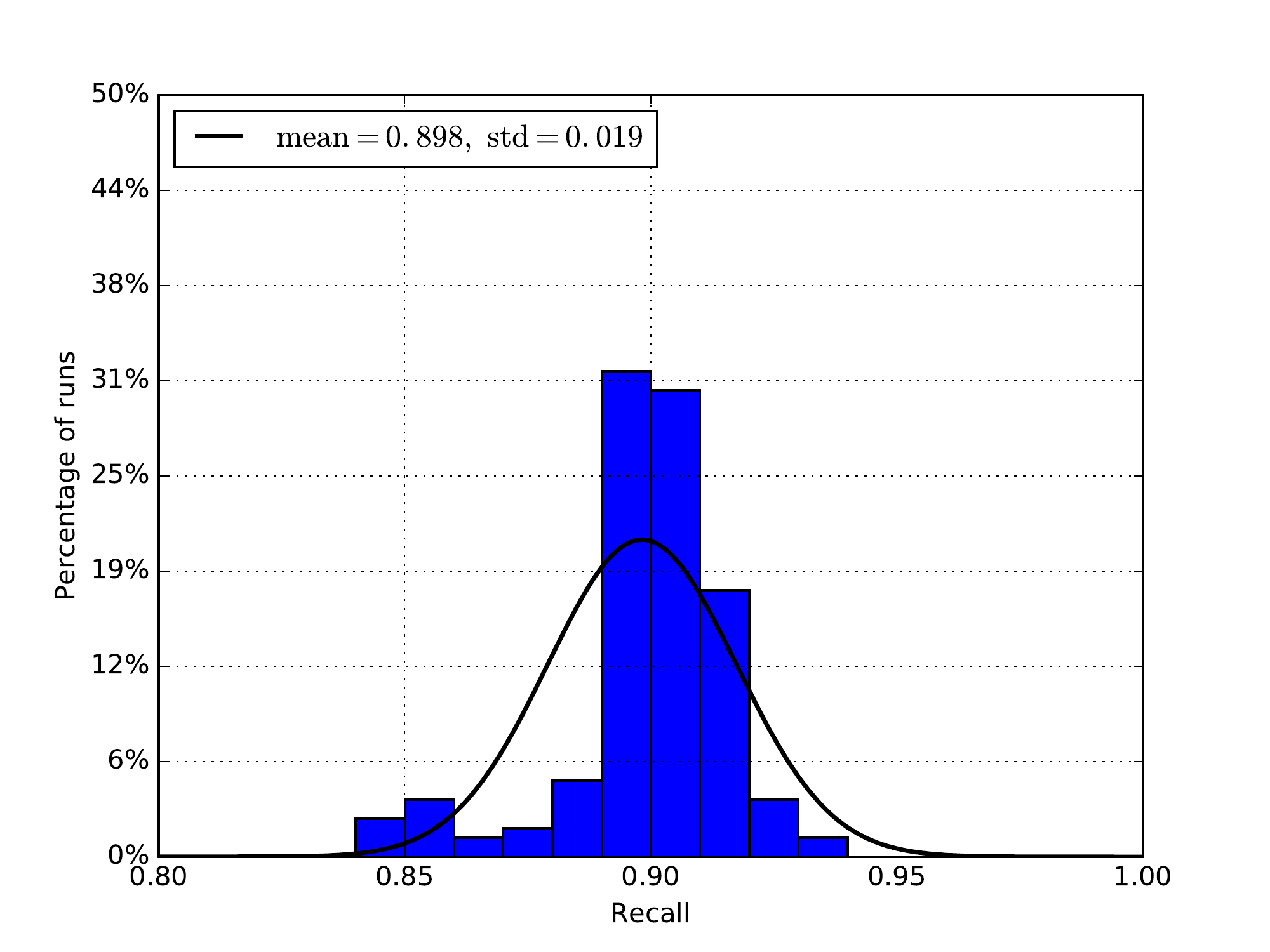}
  \end{center}
  \caption{Karoo GP precision (left) and recall (right) for the magnetometer testing set from 160 runs with varying hyperparameters (tree base depth = 10, maximum tree
  depth = 10, minimum tree depth = 3, population = 300, generations = 100 or 150, tournament size = 10 or 20). The back curves are Gaussian fits to the data. Mean and standard deviation of the Gaussian fits are shown in the legend.}
\label{fig:pr_endxmag}
\end{figure}

\begin{figure}[h]
  \begin{center}
   \includegraphics[width=160mm, height=100mm]{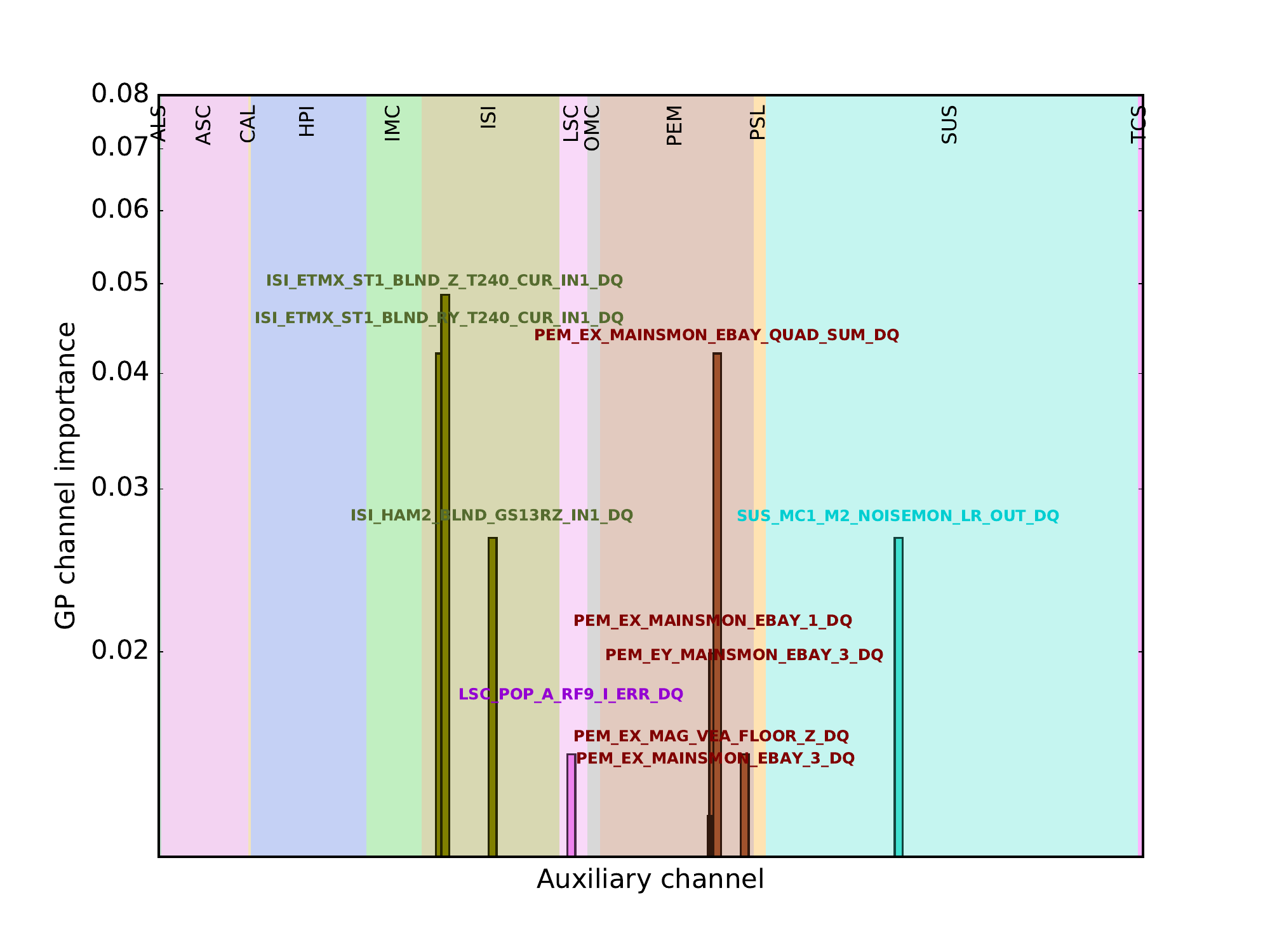}
  \end{center}
  \caption{Channel importance for the magnetometer set from Karoo GP. Only auxiliary channels with GP importance $>0.012$ are shown.}
\label{fig:karoo_magnetometer}
\end{figure}

The above results can be validated with Karoo GP. As the code does not include a standard function to compute the feature importance, we build
a GP analog of this quantity as follows. We run the code multiple times with varying hyperparameters and select the runs that produce a
classification of the testing set with recall and specificity both above a given threshold. Then we define the channel importance by counting
how many times the features of each channel are used in the winning GP multivariate expression of the selected runs and normalize to the total
number of features used. Although this is a rough method of defining the feature importance, it is sufficient for our simple RF validation
task. Figure \ref{fig:karoo_magnetometer} shows the results from eight runs passing a 92\% threshold (out of a total of 160 runs). The
confusion matrix, precision and recall for these runs are shown in Table \ref{table:mag_bestruns}. Histograms of precision and recall for all
160 runs are shown in Fig.\ \ref{fig:pr_endxmag}.

\begin{table}
\begin{center}
  \begin{tabular}{ l | l | l | l | l | l | l }
    {\bf Run}&{\bf TN}&{\bf FP}&{\bf FN}&{\bf TP}&{\bf RC}&{\bf PR}\\ \hline
11&634&42&44&623&0.934&0.937\\
82&659&17&50&617&0.925&0.973\\
91&644&32&50&617&0.925&0.951\\
126&634&42&50&617&0.925&0.936\\
134&638&38&49&618&0.927&0.942\\
146&625&51&45&622&0.933&0.924\\
148&629&47&50&617&0.925&0.929\\
153&644&32&53&614&0.921&0.950\\
  \end{tabular}
  \caption{Confusion matrix for the eight best Karoo GP runs on the magnetometer set (out of 160 total runs) that are used to determine the GP
channel importance. TN=True negatives (background noise correctly identified), FN=False Positives (background noise mis-identified as
magnetometer glitch), FP=False Negatives (glitch mis-identified as background noise), TP=True Positives (glitch correctly identified), RC=
Recall, PR = Precision. Run 11 has the best recall. The run with the best precision (PR=0.998) fails to pass the recall cut-off (RC=0.864) and
is not included in the runs used for the computation of the channel importance.}\label{table:mag_bestruns}
  \end{center}
\end{table}

Because of the limited number of runs used to compute the channel importance, the GP results are generally noisier than the RF results.
However, the GP feature importance is in excellent agreement with the RF importance calculated earlier. Table \ref{table:KarooGP_magnetometer}
and Fig.\ \ref{fig:karoo_magnetometer} show the channels with GP importance larger than 0.012. The auxiliary channels most used by Karoo GP to
separate the magnetometer glitches from the background are two channels of the T240 accelerometer in the EX station
(ISI\_ETMX\_ST1\_BLND\_Z\_T240\_CUR\_IN1\_DQ and ISI\_ETMX\_ST1\_BLND\_RY\_T240\_CUR\_IN1\_DQ) and the PEM\_EX\_MAINSMON\_EBAY\_QUAD\_SUM\_DQ
channel. The first of the ISI channels and the PEM channel are the channels with the highest RF importance for the ISI and PEM subsystem,
respectively. As remarked above, the presence of additional, unrelated channels such as ISI\_HAM2\_\-BLND\_GS13RZ\_IN1\_DQ,
LSC\_POP\_A\_RF9\_I\_ERR\_DQ and SUS\_MC1\_M2\_NOISEMON\_LR\_OUT\_DQ, denotes noisier GP results compared to the RF results. As the GP process
is stochastic, cleaner results may be obtained by increasing the number of runs and/or the efficiency threshold used for the selection of the
winning multivariate expressions. Another way to improve on these results would be to replace the rough count of the channel features with a
better definition of channel importance. 

\begin{table}
\begin{center}
  \begin{tabular}{  l  r  }
    {\bf Auxiliary channel} &{\bf GP Importance}\\ \hline\hline    
    \color{OliveGreen} {\it ISI-ETMX\_ST1\_BLND\_Z\_T240\_CUR\_IN1\_DQ}&\color{OliveGreen} {\it 0.049}\\
    \color{OliveGreen} ISI-ETMX\_ST1\_BLND\_RY\_T240\_CUR\_IN1\_DQ&\color{OliveGreen} 0.042\\
    \color{OliveGreen} ISI-HAM2\_BLND\_GS13RZ\_IN1\_DQ&\color{OliveGreen} 0.027\\
    \color{Violet} LSC-POP\_A\_RF9\_I\_ERR\_DQ&\color{Violet}0.015\\
    \color{RawSienna} {\it PEM-EX\_MAINSMON\_EBAY\_QUAD\_SUM\_DQ}&\color{RawSienna} {\it 0.042}\\
    \color{RawSienna} {\it PEM-EX\_MAINSMON\_EBAY\_1\_DQ}&\color{RawSienna} {\it 0.020}\\
    \color{RawSienna} PEM-EY\_MAINSMON\_EBAY\_3\_DQ&\color{RawSienna} 0.015\\
    \color{RawSienna} PEM-EX\_MAG\_VEA\_FLOOR\_Z\_DQ&\color{RawSienna} 0.013\\
    \color{RawSienna} {\it PEM-EX\_MAINSMON\_EBAY\_3\_DQ}&\color{RawSienna} {\it 0.013}\\
    \color{Turquoise} SUS-MC1\_M2\_NOISEMON\_LR\_OUT\_DQ\color{Turquoise} &\color{Turquoise} 0.027\\
    \hline\hline
  \end{tabular}
  \caption{Auxiliary channels with GP importance larger than 0.012 for the magnetometer set. As in Fig.\ \ref{fig:RF_magnetometer}, different colors denote instrumental and environmental auxiliary channels corresponding to different detector subsystems. Channels in italic denote those selected also by the RF algorithm (see Table \ref{table:RF_magnetometer}).}\label{table:KarooGP_magnetometer}
  \end{center}
\end{table}

\subsection{Air Compressor set}\label{aircompressorresults}

\noindent We repeat the investigation of the previous section for the air compressor set. As this dataset is reduced to only 16 air compressor
noise transients after data preparation, and training occurs only on 2/3 of the glitches (plus an equal number of background triggers) we do
not expect the air compressor results to be as clear-cut as the results for the magnetometer set. However, even if the identification of the
mechanical couplings were to fail, this test would provide important information about the effectiveness of the method for small datasets.
Prompt identification of the origin of mechanical couplings in the detector is of paramount importance for gravitational-wave searches during
observing runs. If the origin of mechanical couplings can be inferred with just a limited number of recorded glitches as soon as they appear
in the detector, the method may prove very useful for commissioning purposes.

\begin{table}
\begin{center}
  \begin{tabular}{  l  r  }
    {\bf Auxiliary channel} &{\bf RF Importance}\\ \hline\hline
    \color{Orchid} ASC-X\_TR\_B\_PIT\_OUT\_DQ&\color{Orchid} .079 \\ 
    \color{Orchid} ASC-X\_TR\_B\_YAW\_OUT\_DQ&\color{Orchid} .169 \\ 
    \color{RoyalBlue} HPI-ETMX\_BLND\_L4C\_RX\_IN1\_DQ&\color{RoyalBlue} .052 \\ 
    \color{RoyalBlue} HPI-ETMX\_BLND\_L4C\_RY\_IN1\_DQ&\color{RoyalBlue} .008 \\ 
    \color{RoyalBlue} HPI-ETMX\_BLND\_L4C\_RZ\_IN1\_DQ&\color{RoyalBlue} .010 \\ 
    \color{RoyalBlue} HPI-ETMX\_BLND\_L4C\_Y\_IN1\_DQ&\color{RoyalBlue} .038 \\ 
    \color{OliveGreen} ISI-GND\_STS\_ETMX\_X\_DQ&\color{OliveGreen} .228 \\ 
    \color{OliveGreen} ISI-GND\_STS\_ETMX\_Y\_DQ&\color{OliveGreen} .012 \\ 
    \color{RawSienna} PEM-EX\_ACC\_BSC9\_ETMX\_Z\_DQ&\color{RawSienna} .055 \\ 
    \color{RawSienna} PEM-EX\_ACC\_EBAY\_FLOOR\_Z\_DQ&\color{RawSienna} .203 \\ 
    \color{RawSienna} PEM-EX\_ACC\_OPLEV\_ETMX\_Y\_DQ&\color{RawSienna} .008 \\ 
    \color{RawSienna} PEM-EX\_ACC\_VEA\_FLOOR\_Z\_DQ&\color{RawSienna} .045 \\ 
    \color{RawSienna} PEM-EX\_SEIS\_VEA\_FLOOR\_Y\_DQ&\color{RawSienna} .024 \\ 
    \color{Turquoise} SUS-ETMX\_L3\_OPLEV\_YAW\_OUT\_DQ&\color{Turquoise} .069 \\ \hline\hline
  \end{tabular}
  \caption{Auxiliary channels with nonzero RF importance for the air compressor set. Different colors denote instrumental and environmental auxiliary channels corresponding to different detector subsystems: Orchid = Alignment Sensing and Control (ASC), royal blue = Hydraulic External Pre-Isolator (HPI), olive green = Internal Seismic Isolation (ISI), sienna = Physical and Environmental Monitor (PEM), turquoise = Suspension (SUS).}\label{table:RF_aircompressor}
  \end{center}
\end{table}

\begin{figure}[h]
  \begin{center}
   \includegraphics[width=160mm, height=100mm]{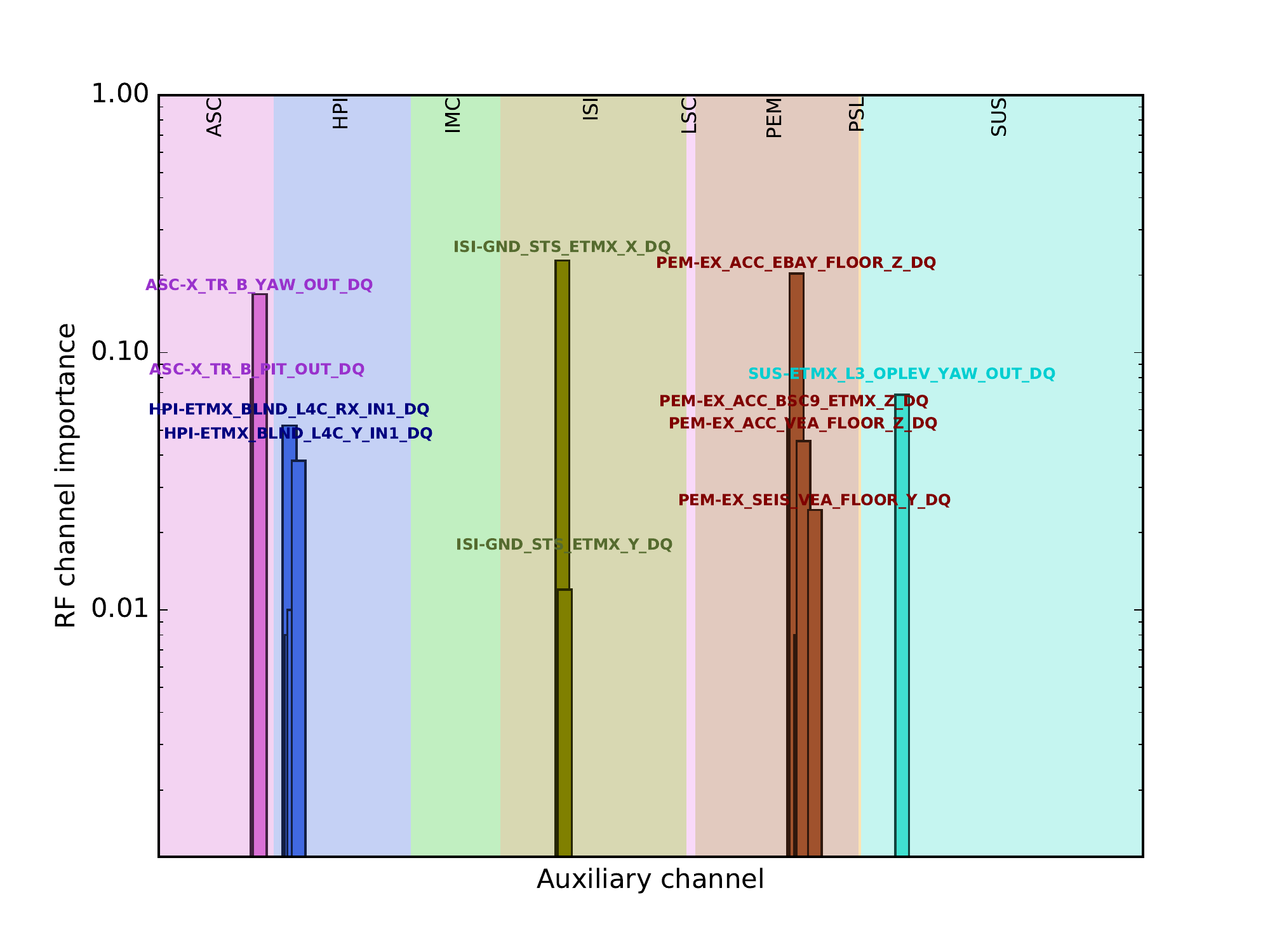}
  \end{center}
  \caption{Histogram of RF channel importance for the air compressor set from the data in Table \ref{table:RF_aircompressor}. Different colors denote different detector subsystems and auxiliary channels: Orchid = Alignment Sensing and Control (ASC), royal blue = Hydraulic External Pre-Isolator (HPI), lime green = Input Mode Cleaner (IMC),
olive green = Internal Seismic Isolation (ISI), violet = Length Sensing and Control (LSC), sienna = Physical and Environmental Monitor (PEM),
orange = Pre-Stabilized Laser (PSL), turquoise = Suspension (SUS).}\label{fig:RF_aircompressor}
\end{figure}

As before, we run the RF code to identify the most relevant auxiliary channels related to the air compressor set. The list of auxiliary
channels with nonzero importance obtained with an RF iteration threshold of 0 is listed in Table \ref{table:RF_aircompressor} and graphically
shown in Fig.\ \ref{fig:RF_aircompressor}. The code identifies five detector subsystems related to the glitch class: the Alignment Sensing and
Control (ASC) subsystem with two auxiliary channels accounting for a total feature importance percentage of $\sim 25$\%, the Hydraulic
External Pre-Isolator (HPI) subsystem with four auxiliary channels accounting for $\sim 10$\%, the Internal Seismic Isolation (ISI) subsystem
with two channels accounting for $\sim 24$\%, the Physical and Environmental Monitor (PEM) with five channels accounting for $\sim 34$\% and
the Suspension (SUS) subsystem accounting for $\sim 7$\%. The winning channels of the HPI, ISI and PEM systems clearly pinpoint the origin of
the glitches as a disturbance originating at the end-test mass X (ETMX). In particular, the PEM signal originates from the accelerometers,
thus signaling a disturbance of seismic origin. The presence of the ASC channels correctly identifies the excess pitch and yaw motion of the
EX test mass due to the seismic coupling. Thus the RF code identifies the same ASC, PEM and ISI channels that were identified by the hveto
pipeline. In addition, the SUS optical lever (OPLEV) channel measuring the yaw (YAW) motion of the bottom stage of the end-test mass X
quadruple suspension (L3), and the HPI channels monitoring the stage of the seismic isolation external to the vacuum between the ground and
the active isolation internal to the chamber clearly show that the disturbance propagates from the ground to the end-test mass through the
various stages of the ETMX suspension. Even with such a limited dataset, the RF algorithm correctly identifies the seismic origin of the
glitches in the EX station. This results seems to indicate that a number of glitches of the order of a few tens, or even less, may be
sufficient to infer the origin of localized disturbances in the instrument and how these disturbances couple to the different detector
subsystems. 

\begin{figure}[h]
  \begin{center}
   \includegraphics[width=160mm, height=100mm]{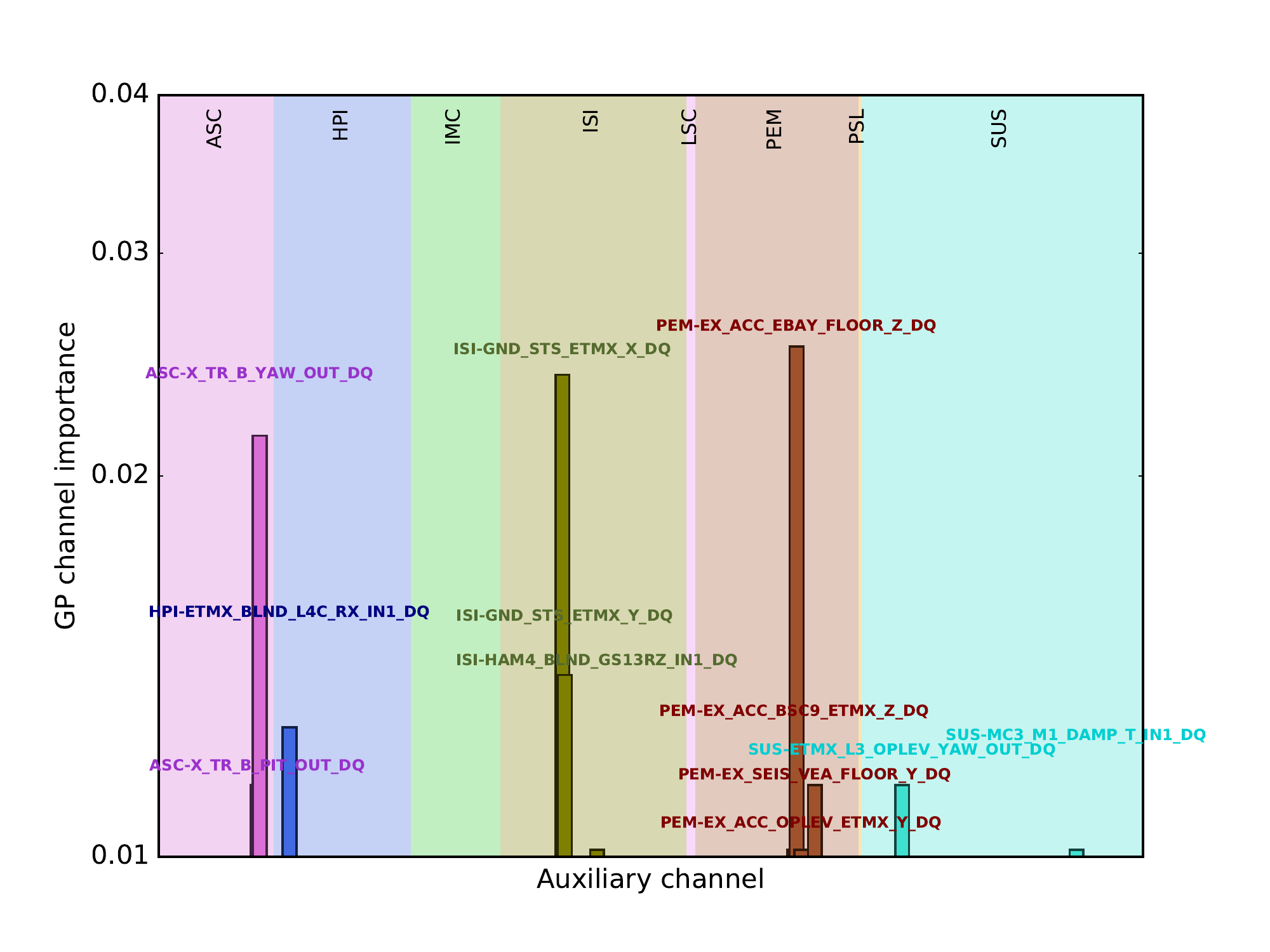}
  \end{center}
  \caption{Channel importance for the air compressor set from Karoo GP. Only auxiliary channels with importance $>0.01$ are shown.}
\label{fig:karoo_aircompressor}
\end{figure}

The above results are validated with Karoo GP by computing the auxiliary channel importance according to the same procedure used for the
magnetometer set. As expected, the Karoo GP results for the air compressor set are less clear-cut than the results for the magnetometer set
due to the limited number of triggers on which the GP algorithm can train. The channels with GP importance larger than 0.01 are shown in Table
\ref{table:karoo_aircompressor} and Fig.\ \ref{fig:karoo_aircompressor} for 34 runs that pass an efficiency threshold on true positives and
true negatives of 66\%. This relatively low threshold is required to build enough statistics for the channel count. As the dataset is limited,
very few runs produce multivariate expression with high true positive/true negative efficiency. Although noisier than the RF results, Karoo GP
results well agree with the RF results. Exceptions are the presence of the ISI-HAM4\_BLND\_GS13RZ\_IN1\_DQ and SUS-MC3\_M1\_DAMP\_T\_IN1\_DQ
channels. These channels are clearly not related to the disturbance and are a noise artifact of the GP algorithm likely to go away with more
runs and a higher threshold on the efficiency for the feature computation. 

\begin{table}
\begin{center}
  \begin{tabular}{  l  r  }
    {\bf Auxiliary channel} &{\bf GP Importance}\\ \hline\hline
    \color{Orchid} {\it ASC-X\_TR\_B\_PIT\_OUT\_DQ}&\color{Orchid} {\it 0.011} \\ 
    \color{Orchid} {\it ASC-X\_TR\_B\_YAW\_OUT\_DQ}&\color{Orchid} {\it 0.022} \\ 
    \color{RoyalBlue} {\it HPI-ETMX\_BLND\_L4C\_RX\_IN1\_DQ}&\color{RoyalBlue} {\it 0.013} \\ 
    \color{OliveGreen} {\it ISI-GND\_STS\_ETMX\_X\_DQ}&\color{OliveGreen} {\it 0.024} \\ 
    \color{OliveGreen} {\it ISI-GND\_STS\_ETMX\_Y\_DQ}&\color{OliveGreen} {\it 0.014} \\
    \color{OliveGreen} ISI-HAM4\_BLND\_GS13RZ\_IN1\_DQ&\color{OliveGreen} 0.010 \\
    \color{RawSienna} {\it PEM-EX\_ACC\_BSC9\_ETMX\_Z\_DQ}&\color{RawSienna} {\it 0.010} \\
    \color{RawSienna} {\it PEM-EX\_ACC\_EBAY\_FLOOR\_Z\_DQ}&\color{RawSienna} {\it 0.025} \\ 
    \color{RawSienna} {\it PEM-EX\_ACC\_OPLEV\_ETMX\_Y\_DQ}&\color{RawSienna} {\it 0.010}\\ 
    \color{RawSienna} {\it PEM-EX\_SEIS\_VEA\_FLOOR\_Y\_DQ}&\color{RawSienna} {\it 0.011} \\ 
    \color{Turquoise} {\it SUS-ETMX\_L3\_OPLEV\_YAW\_OUT\_DQ}&\color{Turquoise} {\it 0.011} \\
    \color{Turquoise} SUS-MC3\_M1\_DAMP\_T\_IN1\_DQ&\color{Turquoise} 0.010 \\ \hline\hline
  \end{tabular}
  \caption{Auxiliary channels with Karoo GP importance above .01 for the air compressor set. Channels in italic denote those selected also by the RF algorithm (see Table \ref{table:RF_aircompressor}).}\label{table:karoo_aircompressor}
  \end{center}
\end{table}

\subsection{Effect of dataset size on algorithm performance}\label{performance}

\begin{figure}[h]
  \begin{center}
   \includegraphics[width=80mm]{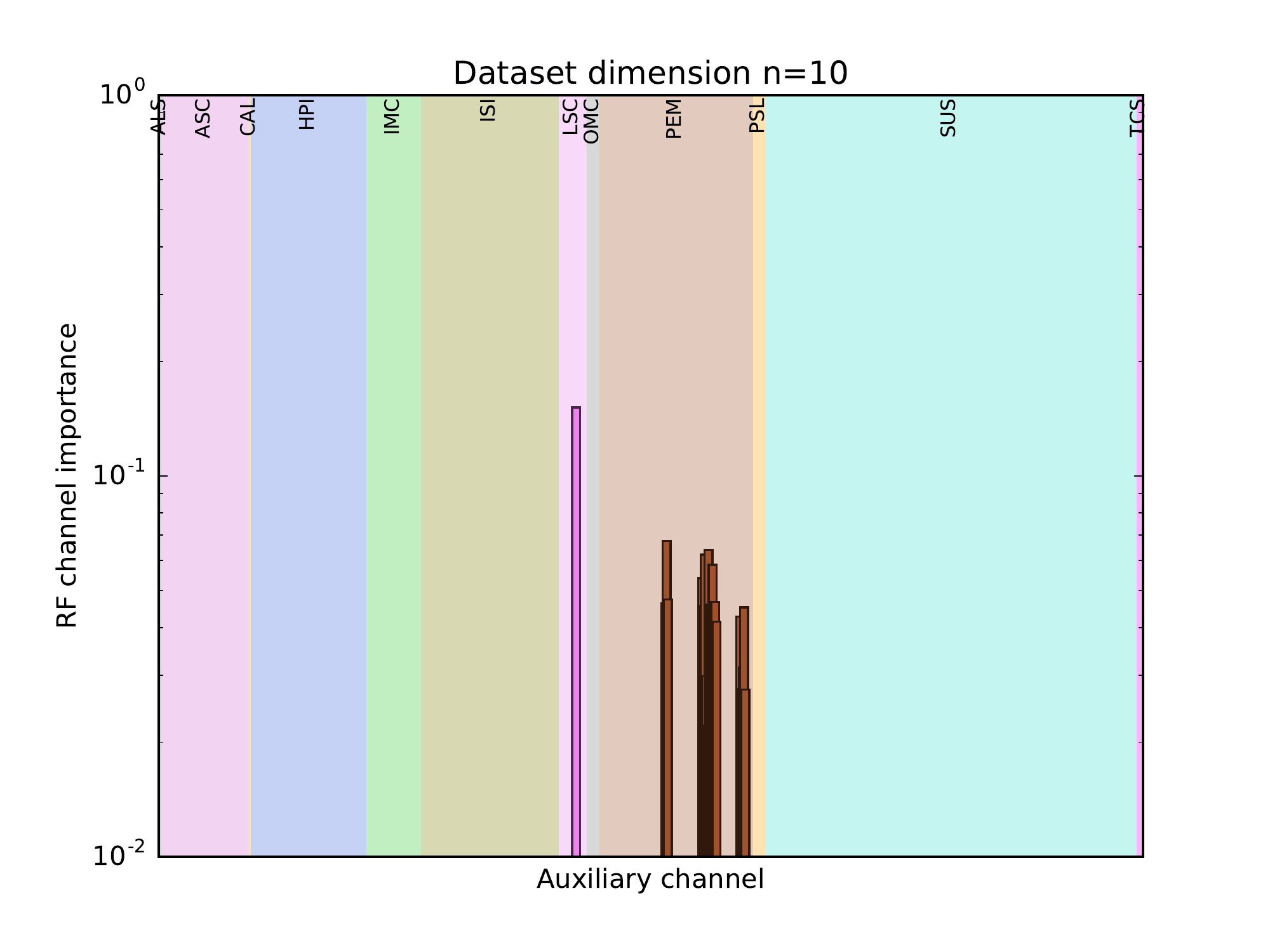}\includegraphics[width=80mm]{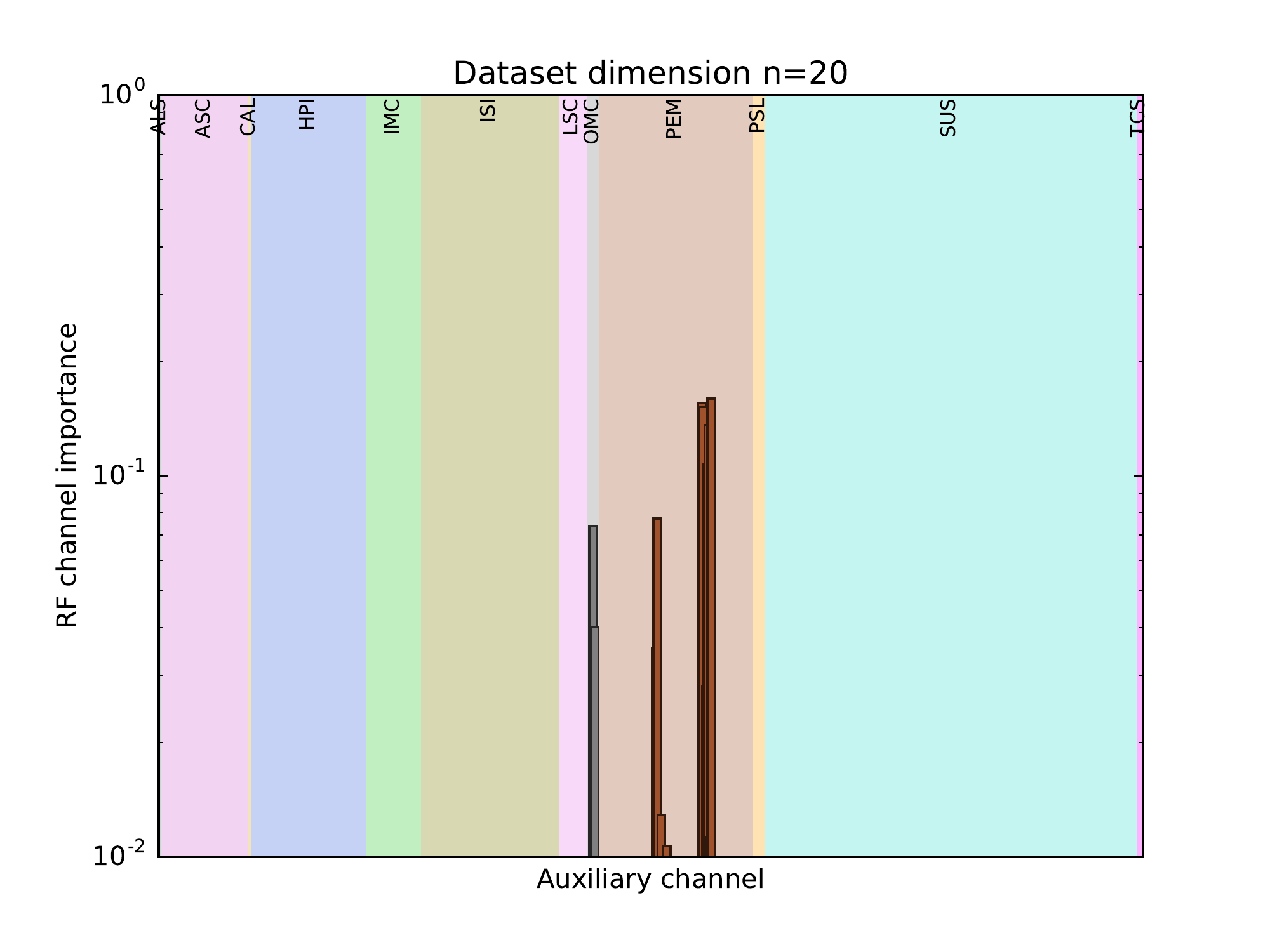}\\
   \includegraphics[width=80mm]{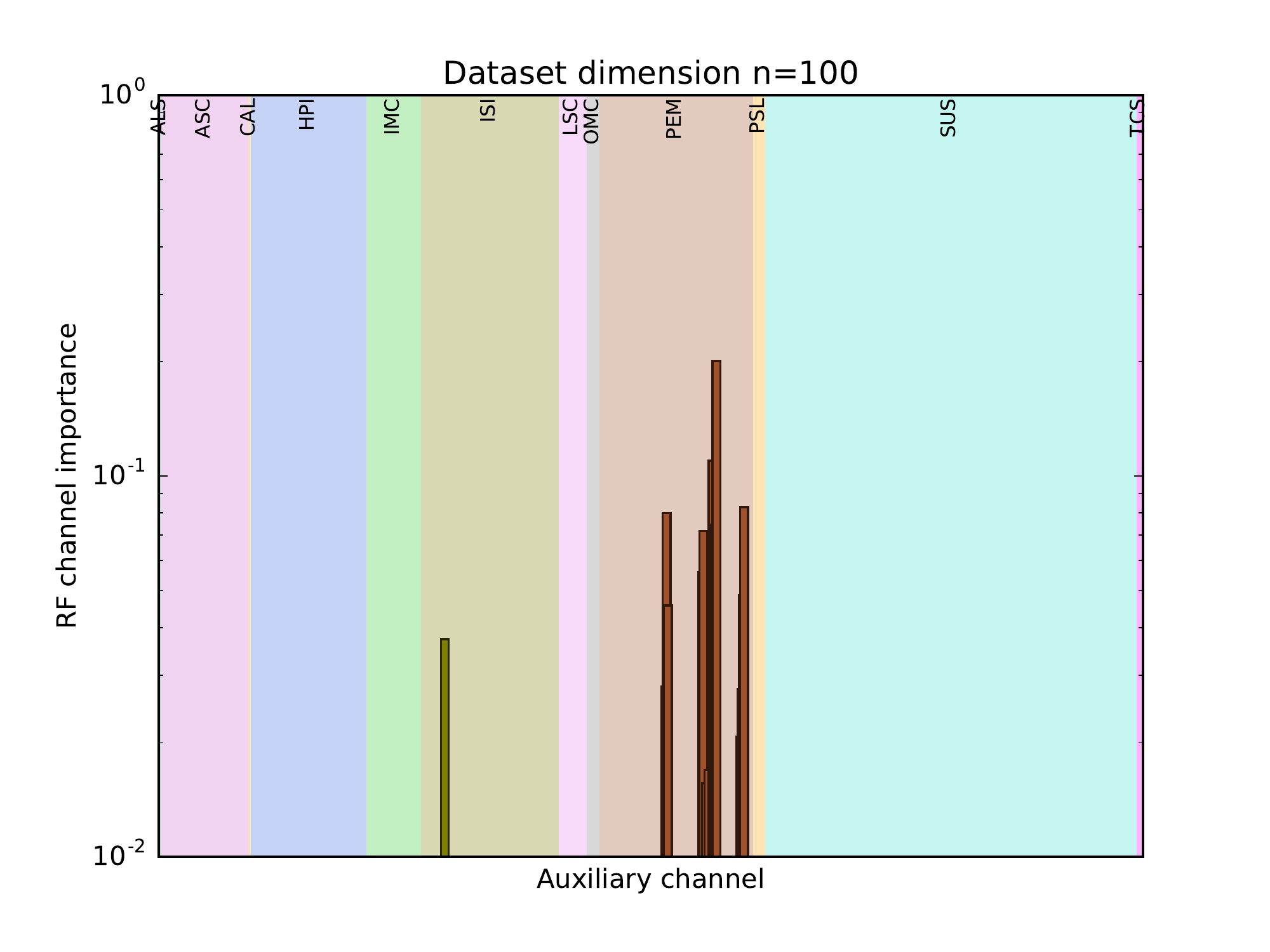}\includegraphics[width=80mm]{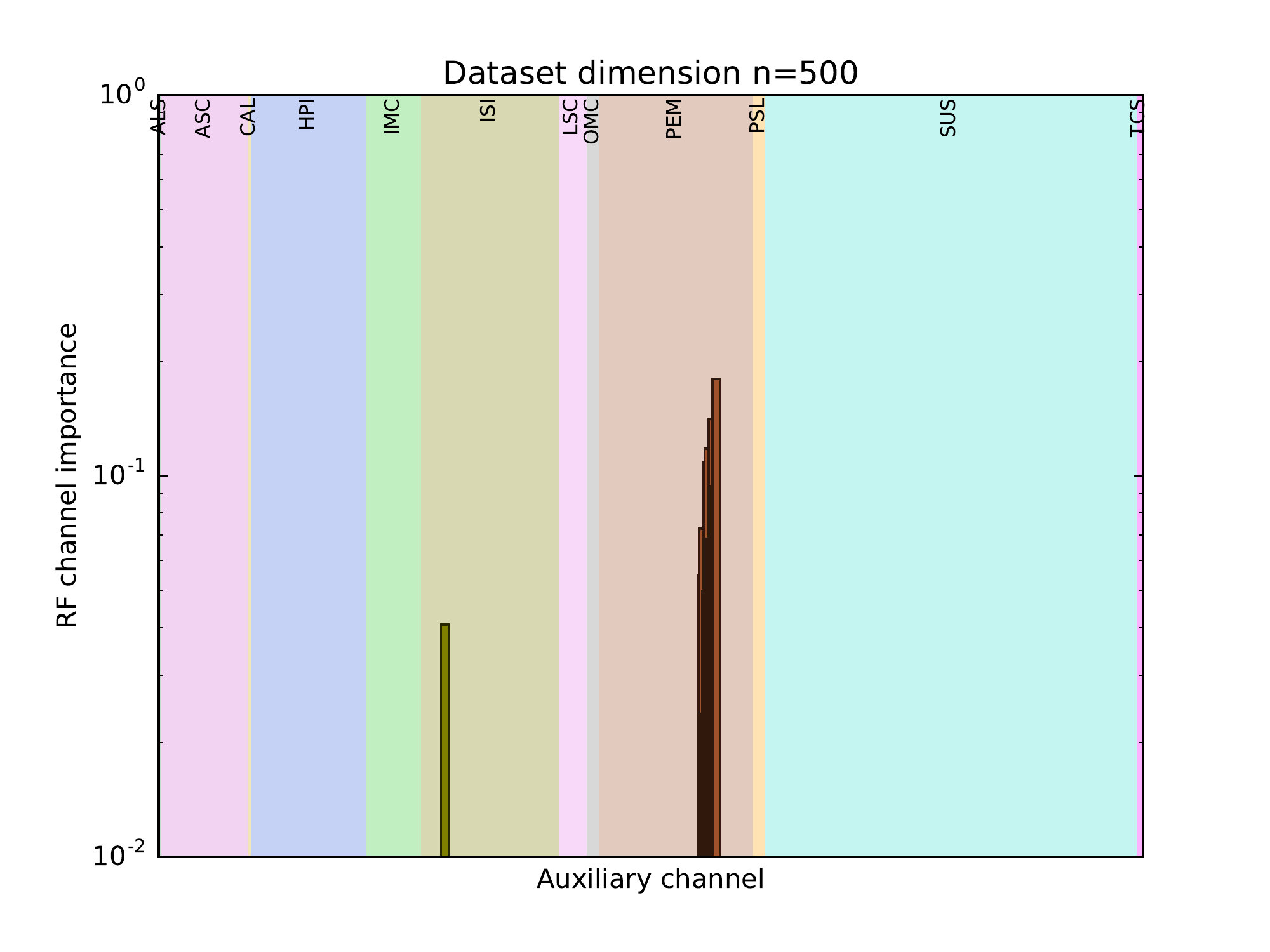}\\   
  \end{center}
  \caption{RF channel importance for dimensionally-reduced magnetometer datasets. From top left, clockwise: $n=10$, $n=20$, $n=500$ and $n=100$.}\label{fig:ROC_magnetometer_RF}
\end{figure}

We have seen in the previous section that even with a handful of glitch times both the RF method and the GP method provide a very good
characterization of the air compressor glitches. In this section we will provide some evidence that small dataset with dimension of a few tens
of glitches can indeed be used to effectively infer the mechanical systems at their origin. In order to do this, we consider again the
magnetometer set and reduce its dimension by selecting a (smaller) number $n$ of magnetometer glitches and building a reduced training dataset
of dimension $2n$ by adding an equal number of randomly selected background glitches. We train the ML algorithms on this $2n$-dimensional set
with fixed identical hyperparameters for all runs and test the result on all the remaining glitches+background that were not used for
training. Figure \ref{fig:ROC_magnetometer_RF} shows the RF results for a number of estimators equal to 500, iteration threshold equal to
0.005 and four different dataset sizes: $n=10$, $n=20$, $n=100$ and $n=500$.

The results for $n=10$ successfully single out the magnetic origin of the glitches in the EX station by identifying the correct PEM EX
channels, although the results are contaminated by the appearance of a few PEM channels of magnetometers located in the interferometer Corner
Station (CS). The importance of these channels stands out only when the dimension of the dataset is strongly reduced ($n=10$, $n=20$, $n=100$)
and may be due to algorithm noise or some contamination in the dataset. For example, some of the magnetometer glitches may be caused by
environmental electromagnetic disturbances which spread-out across the interferometer (e.g., lightning or power glitches) and thus be
erroneously included in the dataset. Likewise, the identification of a Length Sensing and Control (LSC) channel for the $n=10$ dataset and a
couple of Output Mode Cleaner (OMC) channels for the $n=20$ dataset may be due to algorithm noise or indicate that these auxiliary channels are
not safe. Similarly to the CS channels, the importance of these channels stands out only when the dimension of the dataset is strongly reduced.
The results for $n=500$ are basically consistent with the results of the full dataset discussed in Sect.\ \ref{magnetometerresults}.

\begin{figure}[h]
  \begin{center}
   \includegraphics[width=120mm]{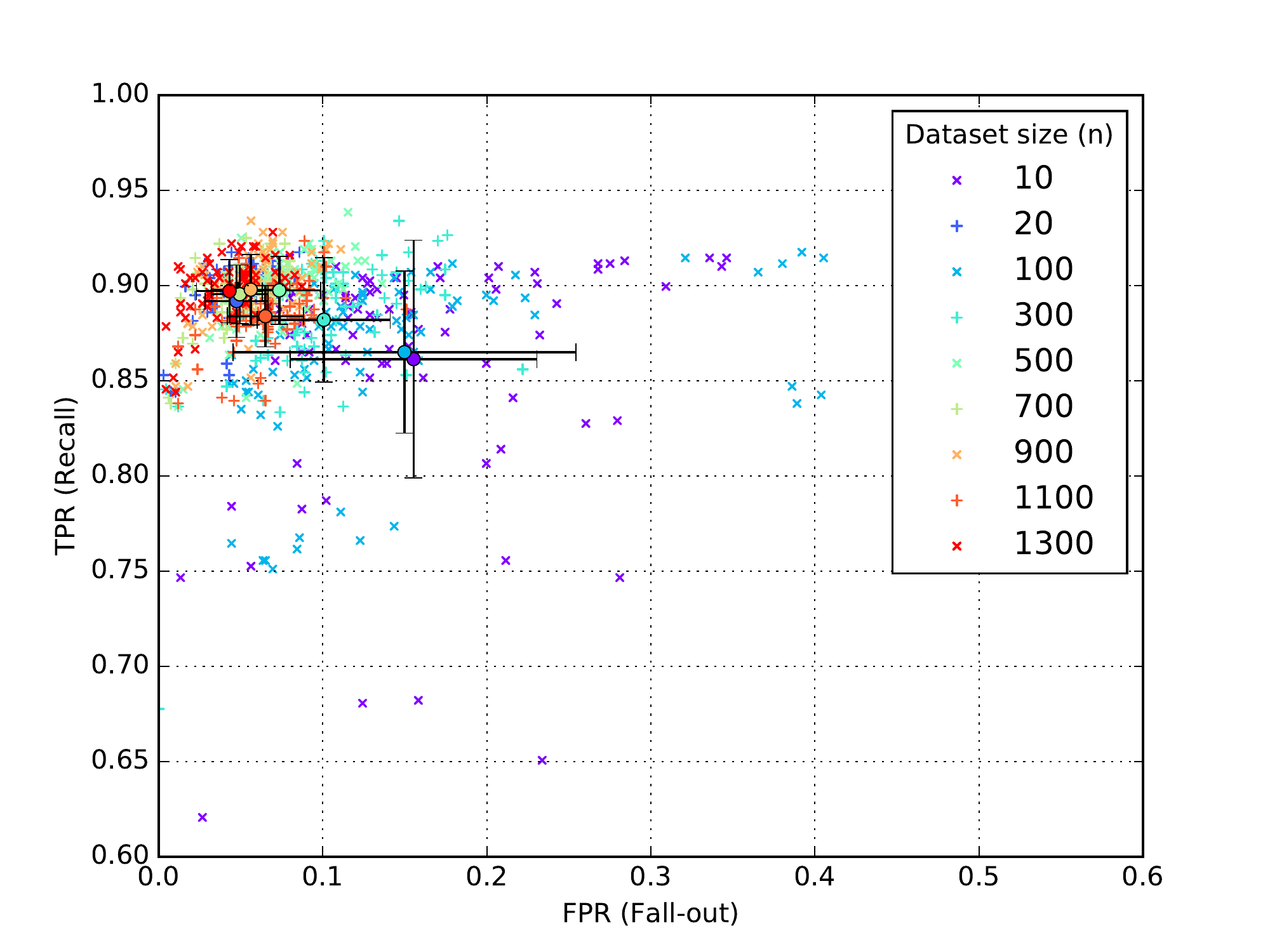}
  \end{center}
  \caption{ROC space for the magnetometer set as the size of training sample is varied. Each point represents the result of a Karoo GP run
with fixed hyperparameters (tree base depth = 5, maximum tree depth = 5, minimum tree depth = 3, population = 300, generations = 100,
tournament size = 20).}\label{fig:ROC_magnetometer}
\end{figure}

\begin{figure}[h]
  \begin{center}
   \includegraphics[width=80mm]{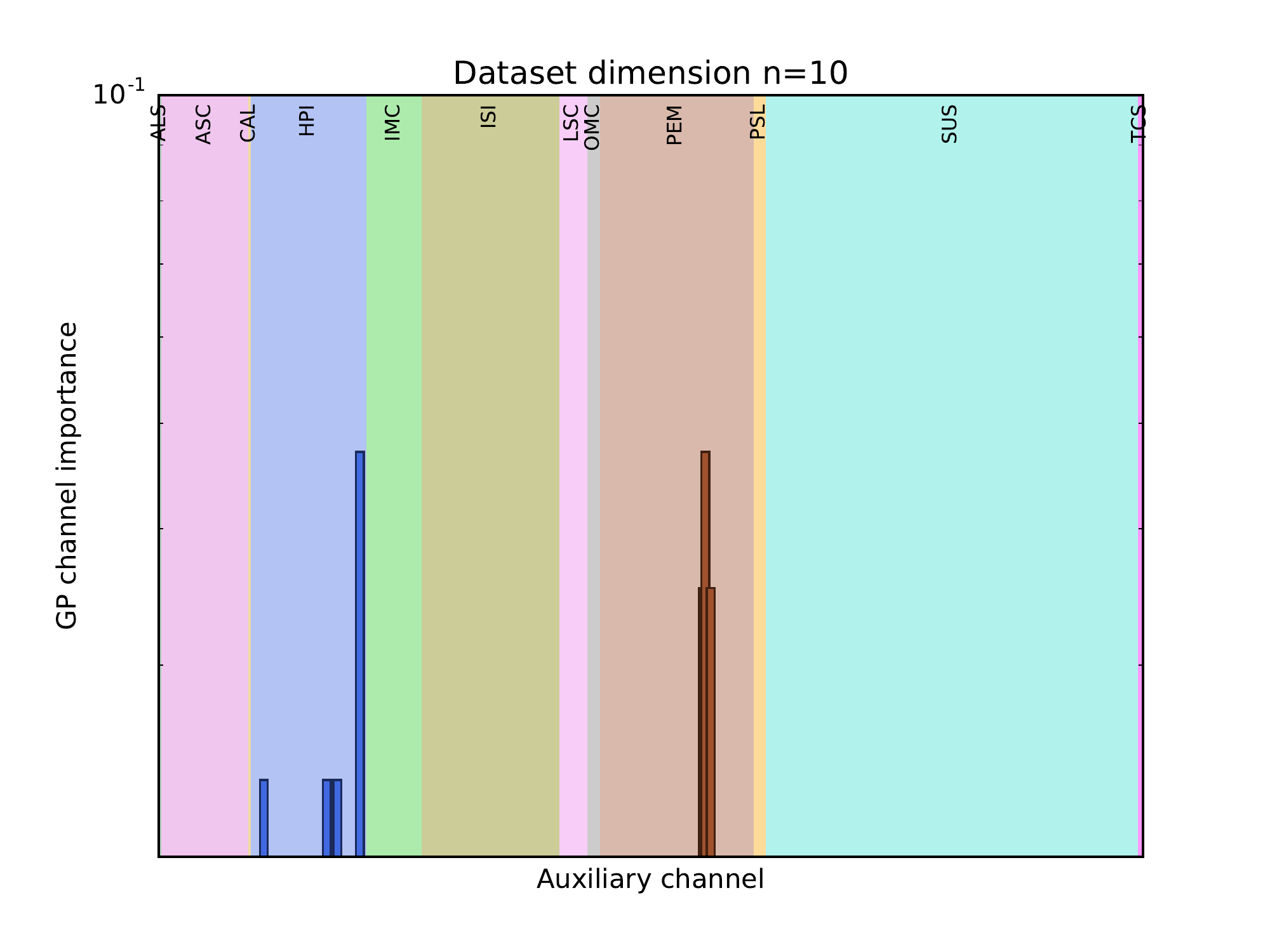}\includegraphics[width=80mm]{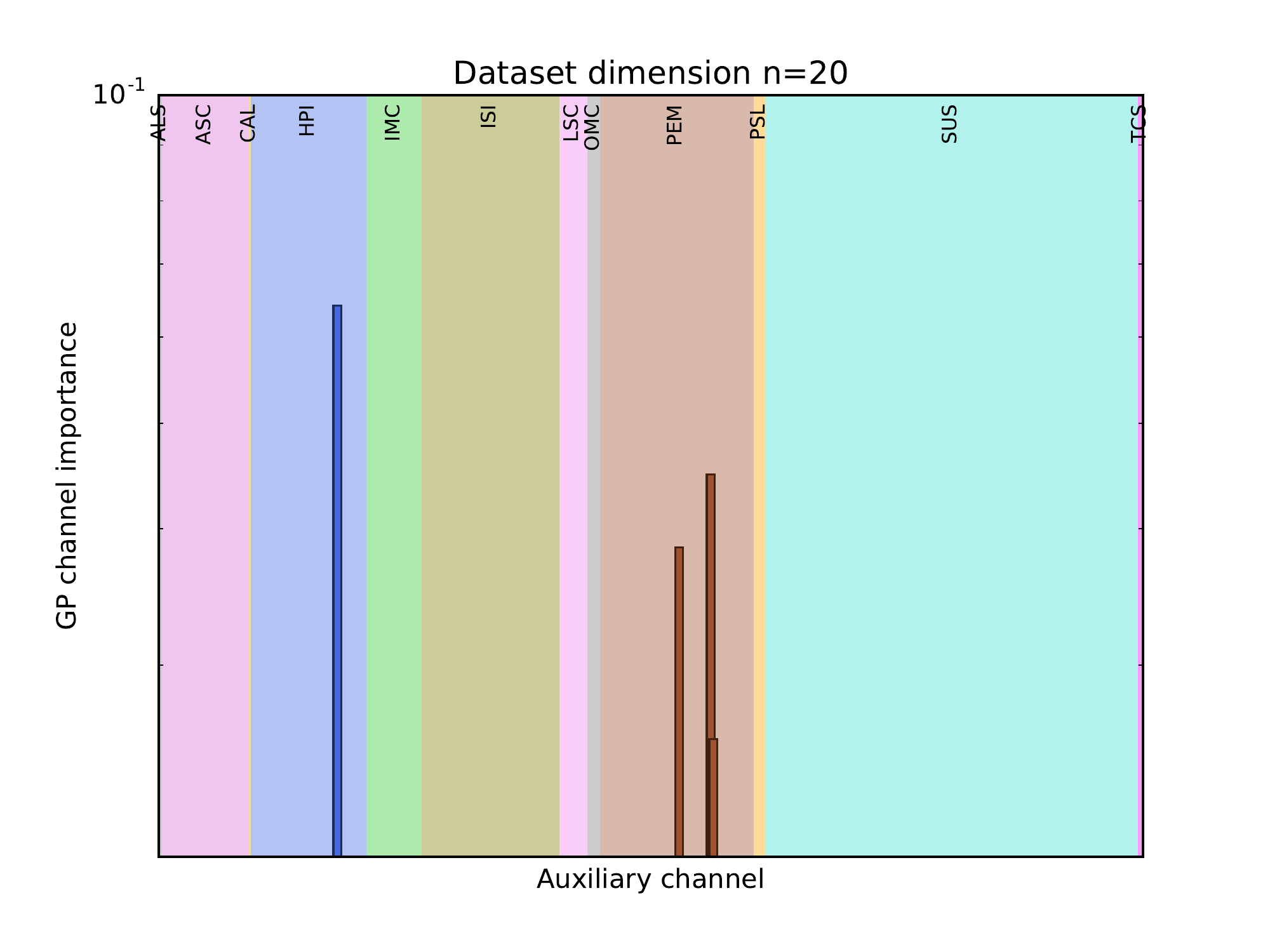}\\
   \includegraphics[width=80mm]{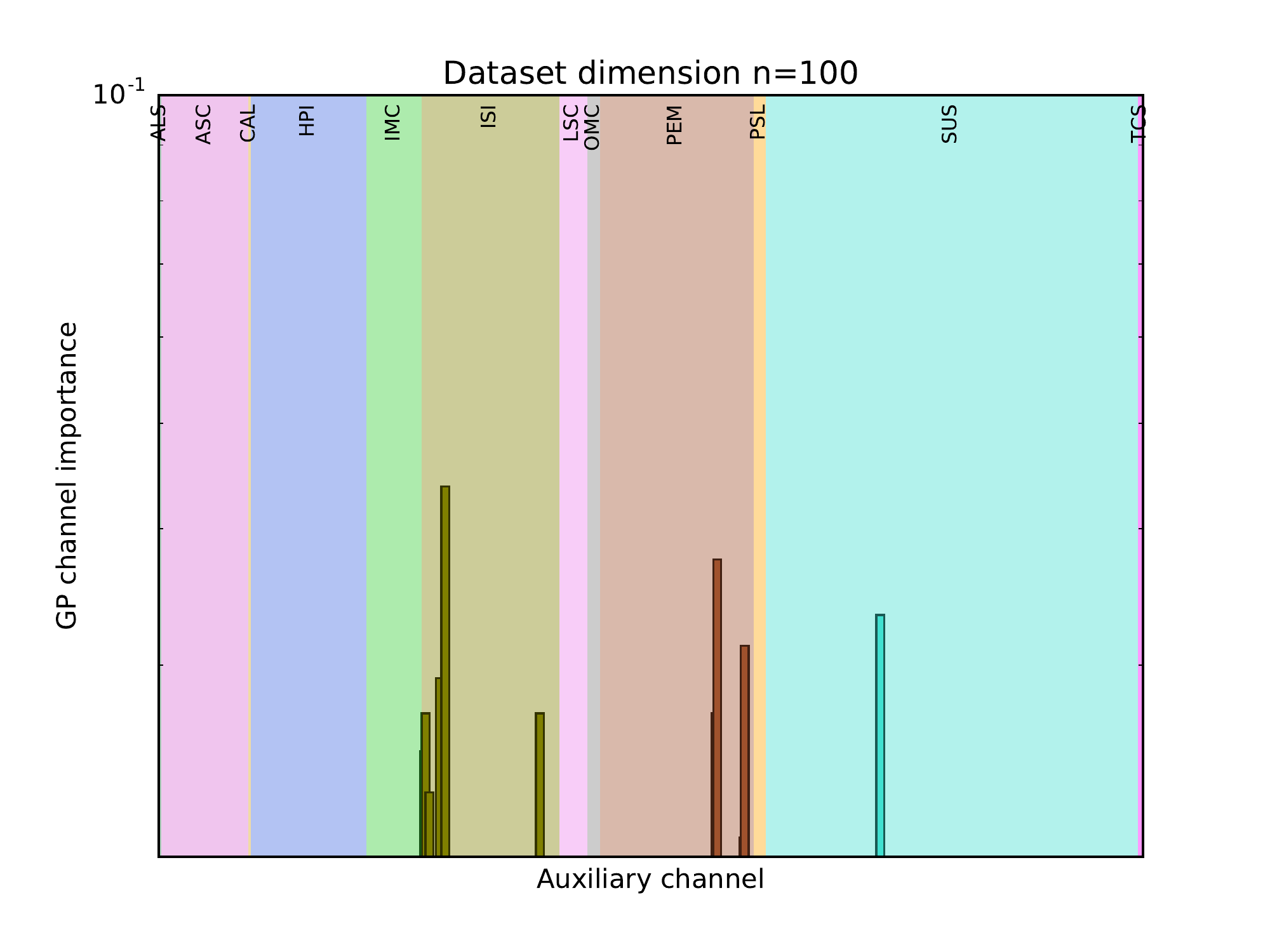}\includegraphics[width=80mm]{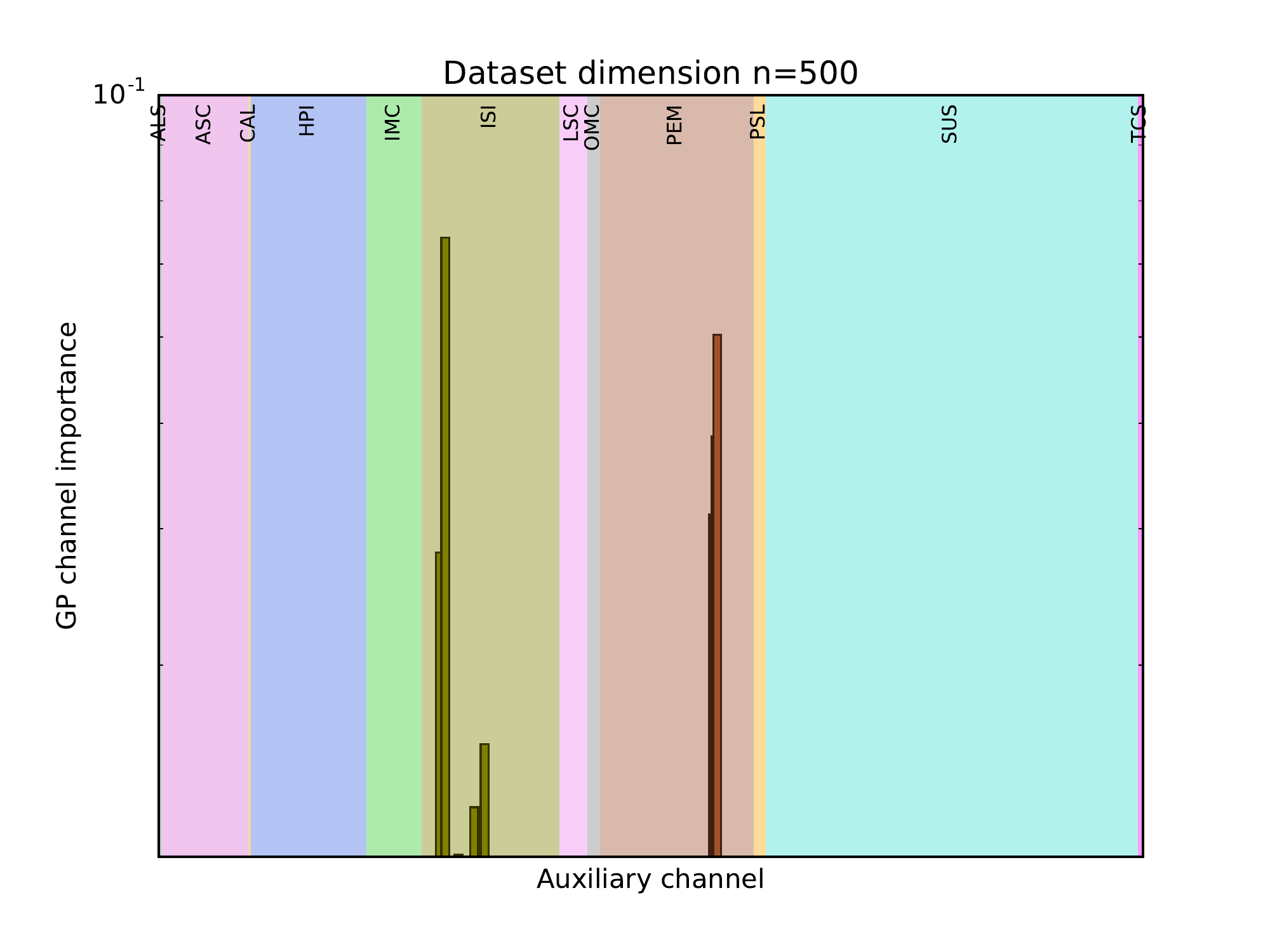}\\   
  \end{center}
  \caption{GP channel importance for dimensionally-reduced magnetometer datasets. From top left, clockwise: $n=10$, $n=20$, $n=500$ and $n=100$. Only channels with importance larger than 0.02 are shown.}\label{fig:ROC_magnetometer_GP}
\end{figure}

We test the accuracy of the GP algorithm as the dataset size is varied by running the code with fixed hyperparameters and then comparing the
confusion matrix of the various datasets. We run Karoo GP 80 times per dataset with tree base depth = 5, maximum tree depth = 5, minimum tree
depth = 3, population = 300, generations = 100 and tournament size = 20. Figure \ref{fig:ROC_magnetometer} shows the ROC space (recall vs.\
fall-out) of the runs with datasets of size $n=10$ to $n=1300$, the latter essentially corresponding to the full dataset. Each point in the
scatterplot represents the result of a Karoo GP run. Average values and standard deviations for each  dataset sample size are also plotted.
Clearly, increasing the dataset sample size improves the binary classification of glitches vs.\ background. However, even for small dataset
size the algorithm performs fairly well, with average recall above $\sim 85$\% and fall-out below $\sim 15$\%. Figure
\ref{fig:ROC_magnetometer_GP} shows the channels with GP importance larger than 0.02 for reduced datasets with dimensions $n=10$, $n=20$,
$n=100$ and $n=500$. The channel importance is extracted from the Karoo GP runs with an efficiency threshold of 88\% for true positives and
true negatives. This lower threshold compared to the threshold used for the full dataset in Sec.\ \ref{magnetometerresults} is required to
obtain enough statistics for the smaller datasets, where Karoo GP performance is worse (9, 15, 32 and 66 runs pass this threshold for the
dimensions $n=10$, $n=20$, $n=100$ and $n=500$ datasets, respectively). As a consequence of the lower threshold, the results are noisier than
for the full dataset. However, the trend is clear as the identification of the correct auxiliary channels improves with the dataset dimension.
The dataset $n=500$ produces a GP channel importance ranking that would undoubtedly allow commissioners to identify the disturbance as
originating in the EX station. Datasets with $n\le 100$ seem to provide a less-clear cut answer. However, as the GP process is stochastic, the
low statistics of these datasets may affect the results. Performing more runs with different hyperparameters is likely to improve the channel
selection. Although less strong than the RF method, the GP method seems to be able to provide useful information on the origin of mechanical
couplings also for datasets with minimal dimensionality.

\section{Conclusions and outlook}\label{conclusions}

\noindent Mitigation of non-astrophysical, instrumental or environmental noise in ground-based interferometric detectors is critical for
improving data quality, reducing the background of gravitational-wave searches and increasing the amount of physical information that can be
extracted from detected signals. Identification of the unwarranted mechanical couplings that cause instrumental noise may greatly help
scientists in performing this effort.

In this paper we have shown that ML methods may be used to infer the possible locations and coupling mechanisms of noise artifacts in LIGO
data. We focused our study on RF and GP algorithms, testing these methods on two sets of non-astrophysical noise transients with known origin
from the first and second observing runs of Advanced LIGO. The {\it magnetometer} dataset contains over 2000 noise artifacts of
electromagnetic nature. The {\it air compressor} dataset contains a few tens of noise artifacts due to environmental seismic coupling. Due to
their well-defined spatial and temporal localization, and their well-understood origin, these datasets provide useful testbeds for the RF and
GP algorithms in two extreme cases of large and small samples.

In order to test the applicability of our methods in real-world situations, we generated ML features derived from the Omicron pipeline, the
standard LIGO-Virgo event trigger generator that is used to identify glitches in LIGO's auxiliary channels during observing runs. The
mechanical couplings at the origin of the disturbances are inferred by computing and then ranking the importance of the Omicron auxiliary
channel features as employed by an ML binary classification scheme of glitches versus background. The RF channel importance is evaluated by
running the standard scikit's RF classifier with fixed number of estimators, and then iterating on the results to remove the features with
importance below a pre-defined threshold.  The GP channel importance is calculated by counting the number of channel occurrences in a subset
of Karoo GP multivariate expressions over a fixed number of runs with varying hyperparameters. 

Both the RF method and the GP method are able to identify the origin of the glitches and infer the relevant mechanical couplings in the
detector. Although the ranking of the auxiliary channels becomes noisier as the size of the training dataset decreases, the algorithms  allow
for a successful identification of the relevant channels even in the case of small datasets with few tens of glitches, such as the air
compressor set. The ability to work with just a handful of triggers is relevant for prompt mitigation of new noise artifacts appearing
suddenly in the data, as it would allow detector commissioners to quickly determine the origin of these artifacts without the need to collect
a large glitch sample.

As our methods rely on standard Omicron triggers which are generated in low-latency at the LIGO and Virgo sites, once a list of GPS times for
a class of (unknown) noise transients with common characteristics in the gravitational-wave strain channel or a given auxiliary channel is
provided, mechanical couplings can be quickly determined. Preparation of a training dataset with a few tens of noise triggers requires a few
minutes on the LIGO computing clusters and channel rankings can be obtained within minutes by either of the two codes. Thus we envision the RF
and GP codes as quick tools for on-demand data quality investigations during observing runs and commissioning periods. 

In addition to providing a new commissioning tool, RF and GP techniques provide the proof of concept that ML can be successfully applied to
the problem of inferring instrumental mechanical couplings. The ML landscape is very rich, with many different methods for scientific data
analysis. It is not unlikely that the results presented in this paper may be further improved by considering other ML approaches, as well as
different methods of feature generation that exploit all aspects of the detector data. The methods presented above could also be easily
adapted to investigations beyond detector characterization. For example, when applied to astrophysical signals, the feature importance ranking
could be used to extract information about the most relevant physical properties of the gravitational-wave sources. Along this direction,
several investigations are currently being pursued in the LIGO and Virgo collaborations to apply ML techniques to different problems from
identification and parameter estimation of gravitational-wave signals \cite{Gabbard:2017lja,Shen:2017jkj, George:2016hay, George:2017pmj} to detector instrumentation \cite{Vajente2017}. The future of ML applied to the analysis of gravitational-wave detector data is certainly bright.

\section{Acknowledgements}\label{acknowledgements}

\noindent This work has been supported by NSF grants PHY-1707668 and PHY-1404139. The authors would like to thank colleagues of the LIGO
Scientific Collaboration and the Virgo Collaboration for their help and useful comments, in particular Dripta Bhattacharjee, Scott Coughlin,
Elena Cuoco, Kate Dooley, Luciano Errico, Hunter Gabbard, Hartmut Grote, Sumeet Kulkarni, Shrobona Loveall, Lorena Maga\~na Zertuche, Kentaro Mogushi, and Jade Powell.

\end{document}